\documentclass[manuscript,screen,authorversion,sigconf,nonacm]{acmart}
\usepackage{url}
\usepackage{hyperref}
\usepackage{enumitem}
\usepackage{makecell}
\newboolean{clean}
\usepackage{multirow}
\setboolean{clean}{false} 

\newcommand{\todo}[1]{%
  \ifthenelse{\boolean{clean}}{}{%
    \textbf{\textcolor{orange}{[TODO: #1]}}%
    {}
  }%
}

\newcommand{\revisit}[1]{%
  \ifthenelse{\boolean{clean}}{}{%
    \textbf{\textcolor{blue}{#1}}%
    {}
  }%
}

\usepackage{kotex}
\AtBeginDocument{%
  }
\usepackage{tabularx}
\usepackage{booktabs}
\usepackage{array}

\setcopyright{none}
\copyrightyear{2018}
\acmYear{2018}
\acmDOI{XXXXXXX.XXXXXXX}
\acmConference[Preprint]{}{}{}
\acmISBN{978-1-4503-XXXX-X/2018/06}

\renewcommand\footnotetextcopyrightpermission[1]{} 

\begin{document}
\title[When the Commons Appropriates a Large Language Model: How WikiVault Reshaped Korean Wikipedia]{When the Commons Appropriates a Large Language Model: \\ How WikiVault Reshaped Korean Wikipedia}



\author{Inhwa Song}
\affiliation{
  \institution{Princeton University}
  \city{Princeton}
  \state{NJ}
  \country{USA}
}
\email{inhwa.song@princeton.edu}

\author{Sohyeon Hwang}
\affiliation{
  \institution{Princeton University}
  \city{Princeton}
  \state{NJ}
  \country{USA}
}
\email{sohyeon@princeton.edu}

\author{Ted Yoo}
\affiliation{
  \institution{Member, Wikimedia Korea}
  \city{Seoul}
  \country{Republic of Korea}
}
\email{ykhwong@daum.net}

\author{Manoel Horta Ribeiro}
\affiliation{
  \institution{Princeton University}
  \city{Princeton}
  \state{NJ}
  \country{USA}
}
\email{manoel@princeton.edu}


\begin{abstract}
Large Language Models (LLMs) have disrupted the balance between content production and quality assurance that sustains knowledge commons, leading many to prohibit or restrict their use.
\textit{But what happens when a community instead appropriates an LLM-powered tool for its own needs?}
We investigate this question through WikiVault, an LLM-powered editing tool developed within the Korean Wikipedia community and used primarily for translation. 
Combining ten interviews, platform-scale analyses, and matched quasi-experimental comparisons, we examine how WikiVault reshaped knowledge production on Korean Wikipedia.
We find the tool 1) drastically amplified the production capacity of a small group of experienced editors, producing longer and more widely viewed articles; 2) shifted work toward reviewing articles and importing content; 3) imported not only content but also editorial judgments from English Wikipedia.
Our findings show how LLM adoption can rebalance the interdependent work that sustains knowledge commons, while illustrating how communities can learn from emerging technologies through use and adapt their governance accordingly.
\end{abstract}



\maketitle

\section{Introduction}

Wikipedia is arguably the world's most consequential digital knowledge commons~\cite{hess_understanding_2007,benkler_wealth_2006,vincent2018examining}, supplying information surfaced by search engines~\cite{vincent2021deeper}, voice assistants~\cite{cohen_wikipedia_2021}, and, most recently, Large Language Models (LLMs)~\cite{miller_new_2025}.
Yet, the largest language editions of the project have taken a largely defensive stance toward using these latter models. 
Newly created policies prohibit LLMs or substantially restrict their use in making contributions~\cite{froneman_failed_2026,wiki_de_kiverbot,wiki_fr_genai,wiki_ai_policies_by_project}, reflecting the uncertainties communities face as LLMs pose substantial threats to the digital knowledge commons they work to build~\cite{froneman_failed_2026,vetter_endangered_2025}.
These commons depend on a fine balance between work that generates new contributions and work associated with governance and quality assurance processes~\cite{ostrom1990governing,frischmann_governing_2014,madison2024knowledge}, both done by a relatively small pool of volunteers~\cite{kittur_he_2007,geiger_when_2013,kittur2007power}. 
In addition to straining servers during data scraping, LLMs threaten to disrupt this balance by allowing contributors to easily produce content that outpaces a community's capacity to review and absorb it~\cite{zhou2025llms,brooks_rise_2024,wiki_aicleanup,wikipedia_g15_2026}.
The appearance of fluency and coherence in the text further renders traditional signals of quality (structure, grammatical fluency, apparent sourcing) essentially moot~\cite{clark2021all,jakesch2023human,walters2023fabrication}.

At the same time, LLM-based tools and affordances may also help address these exact problems. Indeed, the Wikimedia Foundation's recent ``human-first'' AI strategy envisions AI tools supporting contributors from translation and adaptation to moderation and patrolling~\cite{albon_zia_2025_ai_editors,albon2025ai}.
Preliminary research and ongoing Wikimedia initiatives propose and prototype LLM-based systems that can surface factual inconsistencies in Wikipedia for subsequent human review~\cite{semnani_detecting_2025}, help editors identify vandalism~\cite{wikimedia_revision_risk_2023}, and verify whether claims made in Wikipedia articles are indeed supported by their sources~\cite{wikimedia_source_verification_2026}. These explorations recall how Wikipedians have embraced automation more generally in recent decades~\cite{geiger_work_2010,halfaker2012bots,ren_how_2023}, relying on bots and machine learning models to flag potential edits ~\cite{halfaker_ores_2020,smith_keeping_2020} and even perform edits like maintaining templates or reverting vandalism~\cite{zheng_roles_2019,geiger_when_2013}.

However, the emergent nature of LLMs means we have little empirical insight into the impacts of integrating LLM-based affordances into the everyday production and upkeep of digital knowledge commons. 
Existing research has studied how individual editors use general-purpose LLMs~\cite{zhou2025llms} or evaluated novel systems relatively bounded in scope and adoption~\cite{semnani_detecting_2025,wikimedia_source_verification_2026,wikimedia_revision_risk_2023}.
Yet, as Wikipedia language editions navigate consequential decisions about \textit{whether} and \textit{how} to incorporate these technologies~\cite{wiki_ai_policies_by_project,froneman_failed_2026}, understanding their consequences requires studying sustained use across communities with different capacities and needs. 
Smaller Wikipedia language editions, for example, operate with fewer contributors and substantial coverage gaps relative to larger editions~\cite{hale2014multilinguals}.
In such communities, translation from larger editions is an important way to expand coverage~\cite{laxstrom2015content,wulczyn_growing_2016}, a direction that Wikimedia Foundation identified as ripe for support of generative AI~\cite{albon2025ai}. 
Yet, developing such tools responsibly requires understanding not only how to translate text but also how people decide \textit{what} knowledge to bring across and how to adapt it to the linguistic, technical, and editorial context of the receiving community.

\vspace{2mm}
\noindent
\textbf{Present Work.}
Here, we study WikiVault, a rare example of consequential community appropriation of generative AI in Wikipedia.
WikiVault is an LLM-powered editing tool developed and governed by the Korean Wikimedia community that has been available since May 2025.
Although adopted by a relatively small number of contributors, it quickly became a consequential channel for content production, particularly for translating content from other language editions. 
We orient our study towards three research questions:
\begin{enumerate}
    \item[\textbf{RQ1:}] Why and how did the Korean Wikipedia community adopt WikiVault?
    \item[\textbf{RQ2:}] How did WikiVault reshape the work of producing and maintaining content?
    \item[\textbf{RQ3:}] How did WikiVault reshape what knowledge was produced?
\end{enumerate}
We investigate these questions with an integrative, mixed-methods approach that combines ten semi-structured interviews with large-scale quantitative analysis (descriptive and quasi-experimental comparisons) of Korean Wikipedia edit and pageview logs. 

Our findings show that WikiVault substantially expanded Korean Wikipedia's capacity to produce, but also redistributed the work required to sustain that production. 
WikiVault emerged in response to longstanding contributor scarcity and translation burdens and was adopted primarily by experienced editors (\textbf{RQ1}). 
During its first ten months, only 55 editors used the tool, yet WikiVault-assisted contributions accounted for 11.1\% of new articles and 27.0\% of added content.
WikiVault expanded the work that editors could take on, while shifting more of their attention toward reviewing and maintaining the resulting content (\textbf{RQ2}). 
As production began to outpace the community's capacity to review it, editors organized new review efforts and eventually introduced limits on WikiVault use. 
The resulting articles were substantially longer and attracted meaningful reader attention.
Yet WikiVault unevenly changed what was easier to produce. 
By facilitating imports from English Wikipedia, the tool carried over not only information but also editorial choices about what to represent and how.
Editors responded by exercising agency over imported content while also investing in knowledge distinctive to Korean Wikipedia that WikiVault could not readily produce (\textbf{RQ3}).

\vspace{1mm}
\noindent
\textbf{Implications.} 
Our work shows how adopting GenAI-powered tools can reshape the capacity constraints structuring knowledge commons by expanding contributors' capacity to produce, shifting constraints toward maintaining and governing the commons. In particular, WikiVault targeted a specific type of contribution work that was core to Korean Wikipedia: translation. By making knowledge easier to move across languages, the tool thus raised new questions about what knowledge a local commons chooses to adapt versus preserve.

More broadly, as other Wikipedia language editions and knowledge commons grapple with decisions about adopting LLM-powered tools, WikiVault illustrates how communities can learn what place an emerging technology ought to have within their commons through an iterative process of integration and governance. 
Specifically, we identify three conditions that supported this process in Korean Wikipedia---legitimacy, limitations, and legibility---which enabled contributors to introduce the technology in a bounded way, collectively evaluate what emerged through use, and revise practices and constraints in response.

\section{Background and Related Work}
Our study draws on research on how communities sustain the volunteer work required to produce and maintain a knowledge commons, focusing on Wikipedia. 
We first review prior research examining constraints on 
peer production on Wikipedia. 
Then, we review recent work exploring how LLMs might and already do alter the knowledge commons, as well as an older line of research on how technologies in these spaces reorganize work, coordination, and governance. 
Finally, we introduce WikiVault and situate it within Korean Wikipedia as a community-developed LLM-based editing tool that provides a rare opportunity to observe these dynamics in practice.

\subsection{Peer Production on Wikipedia}
Wikipedia is a canonical example of a system created through \emph{commons-based peer production},
i.e., when public goods are assembled through voluntary contributions coordinated outside markets and firms.
Early accounts in the 2000s marveled at the openness of this mode of production, where contributors need not be assigned tasks or formal context, but instead self-select into modular pieces of work~\cite{benkler_coases_2002,benkler_wealth_2006}.
Wikipedia embodies this openness in its technical architecture; with a few exceptions, anyone can edit an article, and edits are published without prior review~\cite{halfaker_rise_2013}.

However, openness alone does not make such systems viable. Work on enduring commons emphasizes that they survive via institutions (boundaries, monitoring, and collectively enforced rules~\cite{ostrom1990governing,hess_understanding_2007,frischmann_governing_2014}) and a long tradition of HCI research has documented the work through which online communities sustain them~\cite{kraut_building_2012,ren_how_2023}.
For example, \citet{kittur_he_2007} show that as Wikipedia grew, a rising share of effort went into \emph{indirect work} like discussion, coordination, and dispute resolution, while \citet{forte_scaling_2008} describe how the difficulty of scaling consensus pushed governance authority outward, decentralizing policy interpretation into local venues such as WikiProjects.
Producing content also generates its own maintenance work: editors must patrol new pages, revert problematic edits, and tag articles for cleanup~\cite{geiger_when_2013,halfaker_dont_2011}.
Much of this labor is easy to overlook. As \citet{star_layers_1999}  observe, the work that holds a system together is often rendered invisible by the very smoothness it
produces; sustaining said smoothness requires the \emph{articulation work} of coordinating what needs repair, who will address it, and when~\cite{schmidt1992taking}.
The maintenance work needed for Wikipedia to flourish draws on a finite and unevenly distributed pool of volunteer capacity.
A relatively small core of highly active editors performs a disproportionate share of Wikipedia’s production and maintenance work~\cite{kittur2007power}, while the standards governing that work are continually negotiated through decentralized policy~\cite{swarts_collaborative_2009}. 

Volunteer capacity can therefore become strained when contributions grow faster than a community can absorb them~\cite{halfaker_rise_2013,halfaker_dont_2011,teblunthuis_revisiting_2018}. 
These capacity challenges are particularly salient across Wikipedia’s hundreds of unevenly sized language editions, which exhibit substantial asymmetries in encyclopedic coverage: topics well developed in one edition may receive limited coverage or be absent in another~\cite{hecht_tower_2010,bao_omnipedia_2012,samir_locating_2024}. 
For smaller editions, translating content from larger editions offers one way to expand coverage despite limited contributor capacity~\cite{wulczyn_growing_2016}. 
Yet translation itself requires substantial labor. Contributors must not only translate text but also adapt it to the linguistic, technical, and editorial conventions of the receiving community. 
Korean Wikipedia, our study site, exemplifies this tension between ambitions to expand coverage and the limited volunteer capacity available to realize them.

\subsection{LLMs and Knowledge Commons}
Generative AI is reshaping how people consume and contribute to knowledge commons. On Stack Overflow, quasi-experimental evidence finds that the release of ChatGPT sharply reduced question volume, raising concerns that generative AI may erode the very corpora on which it depends on~\cite{burtch_consequences_2024,del_rio-chanona_large_2024,shumailov_ai_2024,vetter_endangered_2025}. On Wikipedia, evidence of substitution is more mixed, with heterogeneous topic-level effects on readership and muted effects on editing~\cite{lyu_wikipedia_2025,reeves_exploring_2024}. At the same time, generative AI has entered the production process itself: detection-based estimates find a measurable and growing share of AI-generated text among newly created Wikipedia articles~\cite{brooks_rise_2024}, mirroring findings in scientific writing~\cite{liang_quantifying_2025} and crowd work~\cite{veselovsky_artificial_2023,veselovsky_prevalence_2025}.

Much of the emerging literature has asked whether generative AI outputs meet the epistemic and editorial standards of the communities that receive them. Language models can draft encyclopedic prose when paired with retrieval~\cite{shao_assisting_2024} and support verifiability by locating, checking, and repairing citations~\cite{petroni_improving_2023,semnani_detecting_2025}; a recent systematic review maps this design space for Wikipedia specifically~\cite{reeves_machines_2025}. Yet these capabilities remain uneven in ways that matter for communities with strong editorial conventions. Models asked to apply Wikipedia’s neutrality policy diverge systematically from how editors apply it~\cite{ashkinaze_seeing_2024}. More fundamentally, integrating LLMs can change what contributors are required to do. \citet{zhou2025llms} find that LLMs lower barriers to complex Wikipedia tasks while shifting responsibility for evaluating, verifying, and adapting generated content onto editors. Experienced editors could draw on tacit knowledge of Wikipedia’s norms to perform this work, whereas newcomers were often asked to make consequential editorial judgments before developing those competencies, creating what  \citet{zhou2025llms} describe as a ``paradox of participation.’’

This redistribution of work can alter the relationship between content production and review more broadly. Research on human–AI interaction shows that effective collaboration depends on how tasks are delegated and whether people appropriately calibrate their reliance on AI output~\cite{amershi_guidelines_2019,lai_human-ai_2022,vaccaro_when_2024}. Text outputs that look fluent can complicate this judgment by encouraging over-reliance and reducing the analytic engagement through which errors are caught~\cite{buccinca2021trust}. These concerns have prompted Wikipedia communities to develop rules governing when and how generative AI can be used. German Wikipedia has considered an outright prohibition~\cite{wiki_de_kiverbot}; English Wikipedia has codified narrow conditions for acceptable use alongside an organized cleanup project, and a speedy-deletion criterion for unreviewed LLM-generated pages~\cite{wiki_llm_writing,wiki_signs_ai_writing,wiki_aicleanup,wikipedia_g15_2026}; and policies have developed unevenly across language editions~\cite{wiki_fr_genai,wiki_ai_policies_by_project,froneman_failed_2026}. Alongside these community-level restrictions, the Wikimedia Foundation has articulated a human-first approach to AI, endorsing the automation of routine work while preserving human judgment~\cite{foundation_ai_2025}.

These responses reflect the broader challenge Wikipedia communities face as they decide whether and how to incorporate LLMs into knowledge production. 
The rapid emergence of these technologies has generated substantial controversy as communities weigh their potential against the concerns and any oversight they may require.

\subsection{Technology Adoption in Wikipedia}

Prior research on how Wikipedia communities have incorporated earlier technologies offers some initial insight into how these dynamics might unfold.
Wikipedia's governance is enacted through software as much as through policy.
Bots perform a large share of edits and apply encoded norms at a speed and scale: reverting vandalism and maintaining templates and categories~\cite{geiger_work_2010,halfaker2012bots,zheng_roles_2019} (and even interact in ways their operators did not intend~\cite{tsvetkova2017even}).
These systems can be understood as infrastructure, as they recede into the background of everyday practice and become visible mostly when they break down~\cite{star_steps_1996,jackson201411}.
\citet{geiger_when_2013} demonstrate this empirically, showing that when anti-vandalism bots go offline, damaging edits persist far longer and the community's quality-control work is exposed as the contingent achievement it is.
Machine learning introduced predictive models into this infrastructure.
ORES, for example, exposed predictions about edit and article quality as community-facing services, deliberately designed as \emph{participatory} machine learning so that editors could inspect, contest, and repurpose its scores~\cite{halfaker_ores_2020}.
Article-quality models operationalize existing standards of encyclopedic quality through measurable signals (e.g., article length, structure, sourcing, and revision history), making it amenable to computational assessment~\cite{adler_content-driven_2007,blumenstock_size_2008,warncke-wang_tell_2013,moas_automatic_2023}.

\begin{figure*}[t]
    \centering
    \includegraphics[width=0.8\linewidth, alt={Side-by-side screenshots of WikiVault's translation interface for an English Wikipedia article about the World Nomad Games. The left screenshot shows the source article, with editable English wikitext in the upper panel and its rendered Wikipedia preview below. The right screenshot shows the WikiVault-generated Korean translation, with editable Korean wikitext above and its rendered preview below. The translated wikitext preserves Wikipedia-specific elements such as links, citations, templates, and images. Editors can inspect and edit the generated wikitext and rendered preview before publishing the translated article directly from the interface.}]{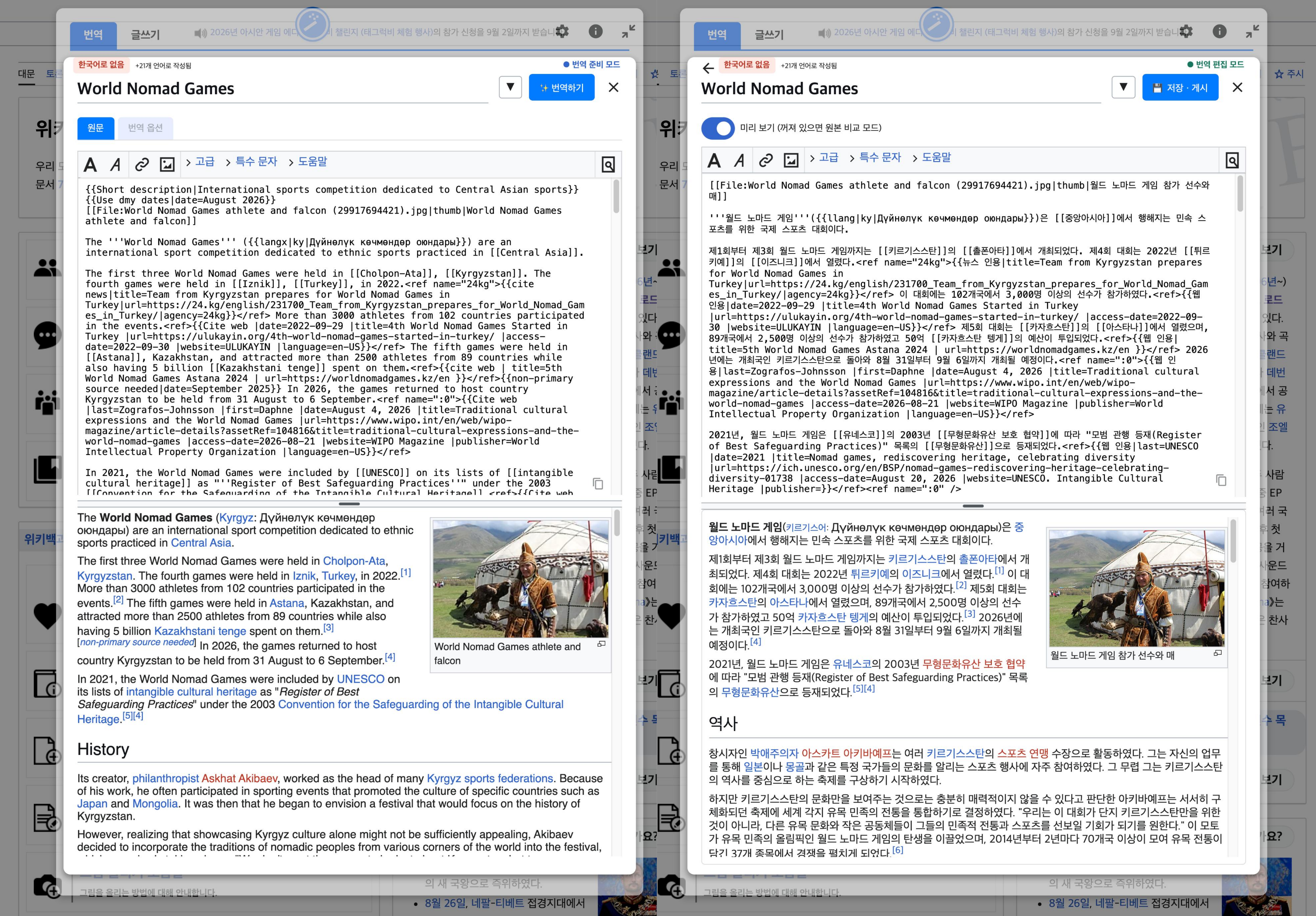}
    \caption{\textbf{WikiVault's translation interface.} As an editor selects an article from English Wikipedia to translate, WikiVault provides wikitext and the rendered preview of the source-language (left), and WikiVault generates Korean wikitext while preserving Wikipedia-specific elements such as links, citations, and templates (right). Editors can edit the wikitext on either side, inspect the rendered Korean preview, and publish the translation directly to Korean Wikipedia.}
    \label{fig:placeholder}
\end{figure*}

However, how software and machine learning algorithms shape Wikipedia depends heavily on how they are incorporated into practice.
HCI has long argued that technologies acquire their consequences through use, as people appropriate, resist, and rework them in ways their designers did not specify~\cite{dourish_appropriation_2003,jackson201411}. 
Past work on social computing highlights several ways these dynamics unfold in practice.
First, automation typically relocates rather than removes work, shifting responsibility onto the volunteers who must adjudicate whatever the system flags or produces~\cite{jhaver_human-machine_2019,schluger_proactive_2022}.
Second, adoption depends not only on technical performance but also on alignment with stakeholder values ~\cite{smith_keeping_2020,chauhan_value_2024}.
Third, design choices around tool integration shape participation: defaults, friction, and how opportunities to contribute are presented can affect both the volume and composition of contributions~\cite{hill_hidden_2021,horta_ribeiro_post_2025}.

Together, this work offers some initial insight into how LLMs \textit{might} reshape Wikipedia, but their consequences cannot be understood from their technical capabilities and design choices alone.
As prior work on technology appropriation emphasizes, technologies acquire their consequences through their incorporation into situated practice~\cite{dourish2001action, dourish_appropriation_2003}.
Understanding what LLMs do to a knowledge commons therefore requires examining what happens when they become part of the the everyday community work.
Yet, doing so has been difficult because LLM use on Wikipedia has often been unsanctioned or restricted and is typically invisible in revision histories, requiring researchers to infer its use and making it difficult to trace assisted contributions after publication. In our work, we examine the case of WikiVault, which provides a rare setting in which these relationships can be observed directly. 


\subsection{Study Context: WikiVault}
\label{sec:study_context}

WikiVault is an AI-assisted editing tool developed within Korean Wikipedia to reduce the repetitive and technical work involved in creating and translating articles. Released in April 2025, the tool automates parts of article production while leaving editors to review and refine its output.
The tool supports two primary workflows:

\begin{itemize}[leftmargin=*]

    \item \textbf{Translation.} Editors can translate an entire English Wikipedia article or selected portions into Korean and provide additional instructions to guide the translation. Because Wikipedia articles contain not only prose but also internal links, citations, infobox templates, and other language-specific wikitext, WikiVault combines LLM-based translation with Wikipedia-specific preprocessing and postprocessing. Before translation, it normalizes the source wikitext and identifies Korean Wikipedia counterparts for links and templates; afterward, it cleans the resulting wikitext and resolves remaining structural issues. The system also accounts for Korean Wikipedia's linguistic conventions, such as using a consistent non-honorific register in encyclopedic prose.
    \item \textbf{Draft generation.} Editors describe in natural language the topic or direction of an article they wish to create. WikiVault uses retrieval-augmented generation (RAG) with Google Search to retrieve relevant sources, generate a short draft grounded in those materials, and incorporate sources into Wikipedia citation templates. The workflow reduces the initial work of locating sources, structuring an article, and formatting citations, while leaving source verification, factual assessment, and further development to the editor.

\end{itemize}

Both workflows are integrated into Wikimedia's broader technical infrastructure. WikiVault is deployed through Wikimedia Cloud Services' Toolforge, which hosts community-developed tools for Wikimedia projects. Edits made through the tool are marked with a \texttt{wikivault} tag in the revision history, making its use visible to other editors and allowing assisted contributions to be identified and traced after publication. Unlike the largely invisible use of general-purpose LLMs, WikiVault therefore operates within Wikimedia's technical ecosystem while leaving an observable record of its use. 

This technical integration reflects WikiVault's broader development within the Korean Wikimedia community. Rather than being introduced as an external AI system, WikiVault was developed by a Korean Wikipedia editor with support from Wikimedia Korea.\footnote{The developer describes WikiVault's motivation, design, and development in a Wikimedia Diff post: \url{https://diff.wikimedia.org/2025/07/15/introducing-wikivault-a-new-chapter-in-wikipedia-contributions-with-ai/}.} The system has continued to evolve through subsequent feature updates and maintenance, making it a case of a community developing and incorporating generative AI into its own editorial practices.

\section{Methods}
We employed an integrative mixed-methods design~\cite{castro2010methodology}, combining computational analyses of platform-scale traces with qualitative inquiry into community governance and editors' situated practices. 
As in \citet{geiger2017operationalizing}, we treated these approaches as complementary ways to investigate how WikiVault reshaped content and its production dynamics within Korean Wikipedia.
We developed the qualitative and quantitative analyses iteratively: insights from community discussions and interviews motivated quantitative analyses of issues raised by editors, while quantitative findings informed subsequent qualitative inquiry.
Given the iterative development of the research, we report quantitative and qualitative findings intertwined in the same sections.

\subsection{Semi-structured Interviews}

\subsubsection {Participant Recruitment and Interviews}
\begin{table*}[t]
\begin{center}
\caption{\textbf{Characteristics of interview participants.}
Participants varied in their Korean Wikipedia editing experience, frequency of WikiVault use, and community permissions, although most were experienced editors.}
\Description{Table summarizing the characteristics of 10 interview participants. Seven participants had at least 10 years of Korean Wikipedia editing experience, one had 3 years, and two had less than 1 year. Five used WikiVault weekly, three monthly, and two a few times. Participants also varied in community permissions, including administrator, autopatroller, and extended confirmed status, while three had no extended permissions.}
\label{tab:participants} 
\begin{tabular}{@{}cllll@{}}
\toprule
\textbf{Alias} & \textbf{Experience} & \textbf{WikiVault use} & \textbf{Extended community permissions} \\
\midrule
P1 & 10 years & Monthly      & Autopatroller, Extended confirmed \\
P2 & 10+ years  & Weekly     & Administrator \\
P3 & 10+ years & Weekly      & None \\
P4 & 10+ years & Weekly      & None \\
P5 & $<$1 year & A few times & Extended confirmed \\
P6 & $<$1 year  & Weekly       & None \\
P7 & 3 years  & Weekly      & Administrator; Extended confirmed \\\
P8 & 10+ years  & Monthly      & Administrator \\\
P9 & 10+ years & A few times     & Autopatroller; Extended confirmed\\
P10 & 10+ years & Monthly     & Autopatroller; Extended confirmed \\
\bottomrule
\end{tabular}
\end{center}

\end{table*}
We conducted semi-structured interviews with 10 of the 55 Korean Wikipedia editors who used WikiVault during our study period.
We recruited participants through a public post on the Korean Wikipedia discussion page (Village Pump\footnote{A central forum for community-wide discussion and coordination, https://ko.wikipedia.org/wiki/위키백과:사랑방}) and through snowball sampling.
In a sign-up questionnaire, participants reported their years of editing on Korean Wikipedia, WikiVault usage frequency, and community permissions.
As shown in Table~\ref{tab:participants}, the sample spanned a range of community permissions and WikiVault usage frequencies.
Participants were mostly experienced editors, which quantitative analyses later found was the norm amongst WikiVault users (see \S\ref{subsec:adoption3}).

The first author, a native Korean speaker, conducted all interviews in Korean over Zoom; interviews lasted approximately 60 minutes. Participants received a 30,000 KRW gift card (approximately USD \$20 at the time) as compensation. We used a semi-structured interview protocol that we iteratively refined throughout data collection to clarify and further pursue themes emerging from earlier interviews and our quantitative analyses. 

Our Institutional Review Board (\textbf{<censored for review>}) approved the study, and all participants provided informed consent before the interviews.
After completing their interview, we invited one participant to join the research team as a co-author because of their extensive knowledge about the community. We made this decision only after their interview, and we retained their interview data as part of the qualitative corpus. Their subsequent involvement brought situated community knowledge into the research process, helping us contextualize our interpretations of WikiVault's use and governance.
To maintain analytical independence, this participant did not code or analyze their own interview, and we triangulated interpretations drawing on their input against other participants' accounts and platform traces.
 
\subsubsection{Data Analysis}
We audio-recorded and transcribed all interviews in Korean. 
To facilitate collaborative analysis among the research team, the first author, fluent in both Korean and English, translated the fully anonymized transcripts into English in chunks using GPT 5.5.
The translation prompt instructed the model to preserve participants' wording, ambiguity, hesitations, and unfinished thoughts without summarizing or interpreting their responses.
The first author then manually reviewed the translations against the original Korean transcripts and revised them where necessary to preserve meaning and context (GPT 5.5 was used solely for transcript translation).

We analyzed the interviews using thematic analysis~\cite {braun2006using}. 
The first author conducted line-by-line coding for all interviews. 
The first two authors then iteratively organized codes into candidate themes using visual code mapping and analytic memos.
The second author, a heritage speaker of Korean, had access to the original Korean transcripts and referred back to them when reviewing translated excerpts and interpretations.
The broader research team discussed candidate themes and supporting excerpts over multiple rounds, refining theme boundaries, relationships, and interpretations. This iterative process produced the final thematic structure used to organize our qualitative findings.
Our qualitative analysis also informed the quantitative investigation; for example, while we initially planned to examine WikiVault's impact on newcomers, interviews and early coding suggested that the tool was used predominantly by experienced editors. We therefore used the platform traces to examine this adoption pattern and shifted subsequent analyses toward the changes in production and review practices raised by participants.
 
\subsection{Quantitative Investigation}

In parallel with our qualitative investigation, we analyzed public Wikipedia traces to characterize how WikiVault’s adoption within Korean Wikipedia and how WikiVault-assisted articles developed following its use. 
Specifically, we used publicly available Wikipedia logs including pageviews, edit histories, and the special tags added to WikiVault-assisted edits. Beyond broad usage trends, we also used two complementary quasi-experimental designs to compare the subsequent trajectories of WikiVault-assisted articles with  comparable ``non-WikiVault'' articles.


\subsubsection{Data Sources}
We constructed our quantitative dataset from publicly available Korean Wikipedia data, combining Wikipedia database dumps\footnote{\url{https://dumps.wikimedia.org/kowiki/}} with Wikimedia pageview records.\footnote{\url{https://dumps.wikimedia.org/other/pageviews/}} 
From the database dumps, we reconstructed article revision histories, including revision timestamps, article length (in bytes), editor identities, and edit metadata. 
We identified bot edits using usernames and bot flags, and reverted revisions using MediaWiki's revert tags. We additionally reconstructed editors' editing histories and community permissions.
We identified WikiVault-assisted revisions through the dedicated tag automatically attached to edits made using WikiVault. This allowed us to identify when individual editors and articles first used WikiVault and to trace their subsequent activity. 
We obtained article topic classifications from Wikimedia's Lift Wing models\footnote{https://wikitech.wikimedia.org/wiki/Machine\_Learning/LiftWing} and identified corresponding English Wikipedia articles through interlanguage links.\footnote{\url{https://www.mediawiki.org/wiki/Interlanguage_links}}
Unless otherwise noted, we restricted our analyses to articles in the `main' namespace, excluded redirect pages, and excluded bot activity.
Our dataset covers July 1st, 2024 - Feb 23rd, 2026 and contains 1,014,325 articles, 63,064 editors, 3,887,073 revisions, and 9,566 WikiVault-assisted revisions.

\subsubsection{Comparing Article Trajectories}
To examine how WikiVault-assisted articles subsequently developed
differently from comparable non-WikiVault articles, we used two complementary matched designs. 
The appropriate design depended on when WikiVault entered the article's lifecycle. 
For existing articles that first received a WikiVault-assisted edit after creation, we could observe pre-treatment trajectories and therefore employed a matched difference-in-differences design.
For articles created using WikiVault, treatment coincided with article creation and no pre-treatment
trajectory existed; we thus employed a matched post-creation comparison corresponding to the post-treatment portion of the same design; c.f. Table~\ref{tab:quasi_design} for a detailed comparison of both designs.

Across designs, we examined article development using weekly article length; readership using weekly pageviews; and editing activity using non-bot edits and content removal.
We modeled pageviews and article length using Poisson regression.
Non-bot editing and content removal, however, were highly sparse, with most article-weeks containing no such activity. 
Given this sparsity, we focused on the extensive margin, using linear probability models to estimate whether WikiVault changed the likelihood that an article received any non-bot edit or had any bytes removed during each week. 
We report additional distributional diagnostics in Table~\ref{tab:outcome_sparsity} in the Appendix.

For \textbf{previously existing articles,} we defined the first WikiVault-assisted revision as the treatment event ($t=0$) and compared article trajectories during the four weeks before and after this event with matched articles that never received a WikiVault-assisted edit.
For each treated article, we constructed a pool of candidate controls that existed at the time of treatment, had never received a WikiVault-assisted edit, and had complete observations during the four pre-treatment weeks ($t=-4,\ldots,-1$). 
We first exact-matched on article creation month and whether a corresponding English Wikipedia article was available at the time of creation.
Within this pool, we performed greedy one-to-one nearest-neighbor matching without replacement using characteristics measured separately in each of the four pre-treatment weeks: pageviews, article length, whether the article received any non-bot edit, and whether any content was removed, along with four-dimensional topic composition scores. 
Pageviews and article length were log-transformed for matching.
We additionally required the absolute difference in log-transformed mean pre-treatment pageviews each week to be no greater than 0.75. This procedure yielded 1,767 matched article pairs, with good covariate balance across behavioral and topical characteristics used for matching (see Appendix~\ref{app:matching-balance}).

Using the matched sample, we estimated an event-study difference-in-differences model using Poisson regression with calendar-week fixed effects. 
For outcomes modeled using Poisson regression, for article $i$ observed in calendar week $t$, we consider:
\[
\log E[Y_{it}]
=
\alpha \, \text{Treated}_i
+
\sum_{k\neq -1}
\beta_k \,
\mathbf{1}
\, \{\text{EventTime}_{it}=k\}
\times  \, \text{Treated}_i
+
\gamma_t
\]
where $\text{Treated}_i$ indicates whether article $i$ received a WikiVault-assisted edit, $\text{EventTime}_{it}$ indexes weeks relative to the treatment event, and $\gamma_t$ denotes calendar-week fixed effects.
For the binary outcomes (editing and content removal), we estimated the corresponding linear probability specification. 
Week $-1$ serves as the omitted reference period. The coefficients $\beta_k$ capture changes in the treated-control difference at each event week relative to their difference immediately before treatment.
The pre-treatment coefficients allow us to assess whether the two groups exhibited differential trends before WikiVault use, while the post-treatment coefficients characterize how their trajectories diverged afterward. 
For Poisson models, we report exponentiated coefficients as incidence rate ratios (IRRs); for linear probability models, we report coefficients as percentage-point differences. 
We cluster standard errors at the matched article-pair level.

For \textbf{articles created using WikiVault,} the same pre-post comparison was not possible because treatment happened with article creation. 
We therefore treated this analysis as the post-treatment portion of the matched design, comparing WikiVault-created articles with matched non-WikiVault articles at the same age after creation. 
For each WikiVault-created article, we constructed a pool of candidate control articles created after WikiVault's deployment that were never edited with WikiVault. Candidate controls had to match the treated article on creation week, have a corresponding English Wikipedia article available at the time of creation, and share the same top macro topic. When multiple eligible controls existed, we randomly selected one control without replacement from the matched stratum. This procedure yielded 4,360 matched article pairs with complete observations over the 12-week post-creation window.
We estimated a separate regression for each of the first 12 weeks after creation, with treatment status as the predictor and calendar-week fixed effects. 
As in the existing-article analysis, we used Poisson regression for pageviews and article length and linear probability models for the binary editing and content-removal outcomes.
The resulting treatment coefficient at each article age captures the difference between WikiVault-created articles and their matched controls, conditional on calendar week. 
Again, we report exponentiated coefficients as incidence rate ratios (IRRs) for Poisson models and percentage-point differences for linear probability models, clustering standard errors at the matched article-pair level.

\section{The Adoption of WikiVault}
Drawing on our qualitative and quantitative data, we trace how and why the Korean Wikipedia community adopted WikiVault (\textbf{RQ1}). 
We found that WikiVault emerged in response to longstanding constraints around editorial capacity, particularly around translation work (\S\ref{subsec:adoption1}).
Its development and rollout built on existing community infrastructure for supporting contributor-developed tools, while incorporating safeguards to make its use reviewable and accountable (\S\ref{subsec:adoption2}).
Finally, adoption remained concentrated among experienced Wikipedians, whose WikiVault-assisted contributions nevertheless accounted for a disproportionate share of new content (\S\ref{subsec:adoption3}).

\subsection{Contributor Scarcity and Translation Burdens Motivated WikiVault}
\label{subsec:adoption1}

\subsubsection{Long-term challenges in Korean Wikipedia: contributor scarcity and translation work}
Participants described WikiVault as emerging in response to longstanding concerns about Korean Wikipedia's editorial capacity and community size (similar to other wikis \cite{halfaker_rise_2013}). 
P9 illustrated the scale of this challenge by contrasting Korean Wikipedia with English Wikipedia, noting that Korean Wikipedia had roughly 10\% as many articles, but only about 2.5\% as many active contributors. 
This limited contributor capacity was compounded by broader social demands on contributors' time; P9 described long-time contributors as aging and increasingly unable to maintain their previous levels of activity as they reached their thirties and forties and became \textit{``tied to their jobs,''} in the societal context of South Korea.

Participants viewed translation as further constraining contributor capacity: they saw it as necessary for expanding the coverage of Korean Wikipedia, but challenging and time-consuming.
P9 explained that the relative scarcity of Korean-language sources made translation from other language editions, especially English Wikipedia, an important pathway to producing content: \textit{``There simply aren't that many other pathways for creating content.''}
Contributors had tried to use various translation tools to aid this work, such as Google Translate and more recently, LLM chatbots. But these tools did not reliably handle wikitext, Wikipedia-specific markup, structure, and jargon.
The Wikimedia Foundation created a ``Content Translation Tool'' in 2014 to support wikitext-aware translation, accounting for markup for internal links, citations, infoboxes, templates, and interwiki links (i.e., links that connect to other wiki projects). 
However, participants like P4, P5, and P8 noted that the tool struggled with wiki-specific elements that differed across language editions and with producing high-quality Korean translations. P1, for example, felt they were \textit{``better off just translating it [themselves] from scratch.''}

\subsubsection{The potential to automate repetitive work}
Participants saw parts of translation work they perceived to be repetitive and unenjoyable as good candidates for automation, motivating efforts to develop new tools to support them. As P2 explained:
\begin{quote}
\textit{``Translation itself takes a tremendous amount of time, but there are parts of it that can be automated.  For people who translate frequently, the same tasks come up over and over---simple grammar corrections, word substitutions, or producing an initial translation. I have thought, why not leave those parts to a program and make translation more convenient?''}
\end{quote}

This view had emerged and shifted over time, with the community having initial apprehension toward LLMs. 
P9 recalled \textit{``an atmosphere of shock and fear''} around ChatGPT within the community, after the tool was released in 2022. However, the persistent burden of translation led P9 to reconsider its potential to help Korean Wikipedia: \textit{``If it can help with translation, does it really matter what kind of tool it is, whether it is LLM?''} 

By late 2024, P10 noted, ideas for AI-assisted editing bots had also begun to surface in public discussions on the Village Pump which highlighted the recurring cost of API use as a constraint.\footnote{https://ko.wikipedia.org/wiki/위키백과:사랑방/2024년\_제50주\#자동AI편집봇이\_필요합니다.} In turn, Wikimedia Korea pointed discussion participants to its existing program to support the development of projects that aided the community.\footnote{https://ko.wikipedia.org/wiki/위키백과:한국\_위키미디어\_협회/활동\_지원/프로젝트\_지원}
Discussions were followed by a more concrete proposal developed between a community member with longstanding technical experience and Wikimedia Korea for an LLM-assisted translation gadget. This proposal became WikiVault.

\subsection{Developing WikiVault and Integrating it Into the Community}
\label{subsec:adoption2}

\subsubsection{The design and development of WikiVault}
WikiVault followed an existing model of community-developed gadgets on Korean Wikipedia.
Participants noted that the Korean Wikipedia community had only a small pool of contributors with the skills to develop technical tools, and that Wikimedia Korea had long supported community members willing to build them.
WikiVault was developed within this arrangement.
Wikimedia Korea provided funding, administrative support, testing, and feedback, while a long-standing contributor familiar with both Korean Wikipedia's editing practices and technical infrastructure led much of the substantive design.

One of our interview participants had been involved in early development.\footnote{We omit this participant's identifier here to avoid linking their statements elsewhere in the paper to their role in WikiVault's development.}
They recalled, \textit{``At the beginning, I had a lot of influence in the design. I thought through much of it myself, so most of my decisions went through as they were.''}
Thus, some decisions reflected their judgments about where such automation would and would not work well based on their experiences as a contributor. 
Although broken templates could technically be removed automatically, they deliberately retained some, as important missing templates might need to be created rather than discarded. 
They also added the drafting feature beyond the originally envisioned translation feature to let editors start articles  \textit{``without relying on translation.''} 
Other decisions addressed recurring burdens they observed in editors' existing workflows, such as removing and restoring references, repairing broken links, and handling other wiki-specific elements before and after translation.

Initial decisions were refined through testing with Wikimedia Korea and other community members. 
Feedback from association staff and editors at offline community meetings\footnote{https://ko.wikipedia.org/wiki/위키백과:오프라인\_모임} identified bugs and problems with translation behavior.
P9 described that the core concern that came up was \textit{``the issue of responsibility,''} explaining,  \textit{``regardless of the tool, the final result has to be the user's responsibility.''} 
WikiVault therefore added a notice stating that \textit{``translation results must be reviewed by the user.''} 
A dedicated WikiVault edit tag was also added at Wikimedia Korea's request \textit{``so that [the edits] could be tracked.''}
Community feedback also informed on how WikiVault would be introduced and used in the community.
Rather than making the tool immediately available to all editors, users had to manually enable it in their preferences.
P9 explained that this \textit{``hurdle''} was intentional, as they worried that otherwise people could quickly create many articles, and leave the community with the resulting cleanup work. 

At this stage, development was not yet announced to the broader community. 
P10 explained that announcing an AI tool before editors could actually try it could lead to a lengthy and \textit{``exhausting and unproductive''} discussion focused on the use of AI itself. 
P10 noted, 
\begin{quote}
    \textit{``I had already seen how, in other Wikipedia communities, simply mentioning AI could generate a lot of discussion. I knew that if we announced that we were developing an AI tool before anyone had even tried it, the discussion would take a long time. So I thought, let's first make the gadget available and try it, and if the response is bad, we can take it down.''}
\end{quote}
Because Wikipedia did not require prior community approval to introduced a gadget, those involved instead favored trying the tool first and removing it if problems emerged. 
As P9 explained, \textit{``It is difficult to find what needs improvement without actually using it.''}

After release, evaluation and feedback broadened to the community at large.
Editors could raise problems and suggest changes through a public discussion space dedicated to the tool.\footnote{https://ko.wikipedia.org/wiki/위키백과토론:도구/WikiVault}
The developer noted, \textit{``After people had a chance to try it, we incorporated various suggestions from users last December, and we’re currently trying to update it based on the feedback we received.''}
As WikiVault became used more widely, these discussions came to concern not only the tool's design but also how its use should be governed, which we return to in \S\ref{sec:reorganizing_workflow}.

\begin{table}[t!]
\centering
\small
\caption{\textbf{WikiVault's contribution to Korean Wikipedia during the first nine months after deployment.}}
\Description{Table summarizing WikiVault's contribution to Korean Wikipedia during the first nine months after deployment. WikiVault was used by 55 active editors, 0.11\% of all active editors, and accounted for 9,566 edits, or 0.63\% of all edits. Despite this limited adoption, WikiVault accounted for 7,607 newly created pages, or 11.10\% of all new pages, and 220 million added bytes, or 27.00\% of all added content. Pages associated with WikiVault received 3.23 million pageviews, representing 36.6\% of the pageviews reported in the table.}
\label{tab:descriptive}
\begin{tabular}{lrrr}
\toprule
\textbf{Metric} & \textbf{Total} & \textbf{WikiVault} & \textbf{\% WV} \\
\midrule
Pages created                 & 68,642      & 7,607        & 11.10\% \\
Edits                         & 1,513,717   & 9,566        & 0.63\% \\
Added bytes                   & 813,680,042 & 220,028,216  & 27.00\% \\
Active editors ($\geq$1 edit)        & 49,080      & 55           & 0.11\% \\
Avg monthly act. edit. ($\geq$1 edit) & 5,761.5 & 20.5 & 0.36\% \\
Pageviews  & 8,827,954 & 3,231,646 & 36.6\% \\
\bottomrule
\end{tabular}
\end{table}
\begin{table}[t]
\centering
\small
\caption{\textbf{Distribution of WikiVault edits across namespaces.}}
\Description{Table showing the distribution of WikiVault edits across four namespaces. Most WikiVault edits were to encyclopedia articles in the Main namespace, with 9,566 edits, compared with 1,563 Template edits, 343 Category edits, and 151 User edits.}
\label{tab:wv_namespace}
\begin{tabular}{llr}
\toprule
\textbf{Namespace} & \textbf{Purpose} & \textbf{WikiVault Edits} \\
\midrule
Main      & Encyclopedia articles           & 9,566 \\
Template  & Formatting and citation templates & 1,563 \\
Category  & Article categorization          & 343 \\
User      & Personal drafts / sandboxes     & 151 \\
\bottomrule
\end{tabular}
\end{table}

\subsubsection{Community mediated adoption of WikiVault}
Since the developer team had deliberately made it so that users had to go to their own settings and activate WikiVault manually, any potential user had to first be aware of the existence of WikiVault. 
WikiVault was initially introduced through Village Pump and community social channels, but some participants first encountered WikiVault through offline meetings (P7), edit-a-thons\footnote{https://ko.wikipedia.org/wiki/위키백과:에디터톤} (P5), and word-of-mouth from other editors (P6). For example, P5 learned to use WikiVault at an offline neurodiversity edit-a-thon he regularly attended, where a senior editor introduced the tool and demonstrated how to use it.

Community events and programming created structure that helped newcomer editors learn how to use WikiVault and begin contributing with it.
P6, a newcomer of Korean Wikipedia, was a particularly illustrative case.
P6 initially used Wikipedia's default translation tool\footnote{i.e., the Content Translation Tool created in 2014: https://en.wikipedia.org/wiki/Wikipedia:Content\_translation\_tool} because he did not yet know about WikiVault.
While discussing mistakes he had made editing an article with his assigned mentor,\footnote{https://ko.wikipedia.org/wiki/위키백과:새\_사용자\_경험/새\_사용자\_멘토} \textit{``they told me that if I was going to translate, it would be better to use WikiVault. I learned to use WikiVault from them, and I've been using it ever since.''}

\begin{figure*}[t]
    \includegraphics[width=0.9\linewidth, alt={Four-panel figure showing weekly activity on Korean Wikipedia before and after WikiVault's deployment, with separate lines for overall and WikiVault-associated activity. The panels show active editors, pageviews, edits, and added bytes. WikiVault activity begins after deployment and grows over the following weeks. The number of WikiVault editors and edits remains small relative to overall activity, while WikiVault-associated pageviews and especially added bytes constitute more substantial amounts. WikiVault-associated pageviews and added bytes rise sharply during the first several months after deployment and generally decline afterward. Solid lines represent overall activity, dashed lines represent WikiVault-associated activity, and vertical lines mark deployment.}]{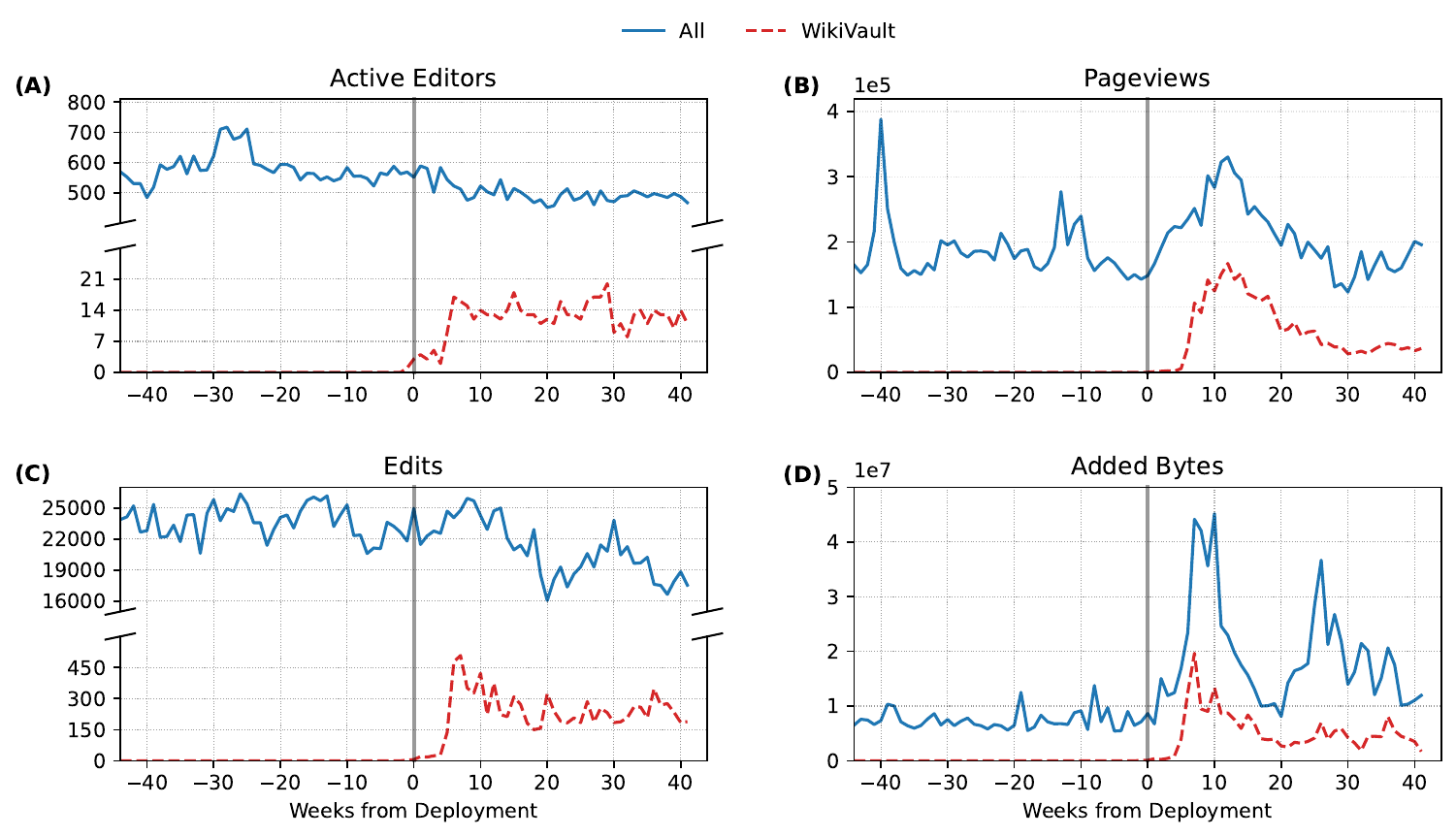}
    \caption{\textbf{Weekly activity before and after WikiVault's deployment.}
Each panel shows weekly activity across Korean Wikipedia (solid) and activity associated with WikiVault (dashed), including active editors, pageviews, edits, and added bytes. The vertical line marks WikiVault's deployment on May 1, 2025. We note that WikiVault has enabled a very small fraction of the editors to contribute a disproportionally large amount of content to Korean Wikipedia.}
    \label{fig:descriptive}
\end{figure*}

We note that the intentional \textit{``hurdle''} designed to constrain access to WikiVault also had consequences.
P2 commented that editors who did not follow the Village Pump or broader community spaces, instead \textit{``quietly editing''} individual articles, could have difficulty learning about WikiVault and finding and enabling it in their preferences. Said otherwise, people already embedded or actively engaged with the community were far more likely to learn about WikiVault and start using it.

\subsection{WikiVault Adoption Was Concentrated but Consequential} \label{subsec:adoption3}

WikiVault adoption ultimately remained concentrated among a relatively small number of editors, but their use of the tool became consequential for content production.

As shown in Table~\ref{tab:descriptive}, only 55 editors adopted WikiVault during the first ten months following deployment, representing 0.12\% of all editors who contributed during the observation period. 
On average, 20.5 editors used WikiVault each month, representing 0.36\% of monthly active editors. 
Nevertheless, these contributors created 7,607 new articles---11.10\% of all newly created pages---and contributed to over 220 million added bytes---27.00\% of all added content---while accounting for only 0.63\% of all edits. 
Pages edited with WikiVault also accounted for 36.6\% of all pageviews during the observation period.
Although our analyses focus on WikiVault edits in the main article namespace\footnote{A namespace is a collection of pages that falls under a common type, such as Main/Article, User, Draft, etc.} (9,566 edits), we also observed use in supporting namespaces such as templates (1,563 edits) and categories (343 edits; see Table~\ref{tab:wv_namespace}).

WikiVault's translation feature accounted for nearly all observed WikiVault use. 
Among WikiVault edits from July 1st, 2025 onward, when the use of the translation feature could be reliably identified from edit summaries, 98.6\% of edits to existing articles and 99.3\% of article-creation edits were identifiable as uses of WikiVault's translation feature (compared with 1.4\% and 0.7\%, respectively, for the drafting feature).
Thus, although WikiVault also included a drafting feature, this feature accounted for a very small fraction of observed use.
Adoption was concentrated among experienced editors. Nearly 80\% of WikiVault users made their first WikiVault edit only after completing more than 1,000 edits. 

Figure~\ref{tab:descriptive} shows how this concentrated but consequential use unfolded over time.
WikiVault activity remained minimal during the first few weeks following deployment before increasing steadily approximately one month later. 
The number of active WikiVault editors then stabilized at roughly 10-20 per week, remaining small relative to the roughly 500 active editors overall (Figure~\ref{fig:descriptive}A).
WikiVault edits similarly constituted a small fraction of weekly editing activity throughout the observation period (Figure~\ref{fig:descriptive}C)
Yet this relatively limited editing activity corresponded to much larger shares of added bytes and pageviews.
WikiVault edits contributed to a substantial share of weekly added bytes, particularly during the first several months after deployment (Figure~\ref{fig:descriptive}D). Pages edited by WikiVault also received a substantial share of weekly pageviews (Figure~\ref{fig:descriptive}B)

Taken together, these patterns show that WikiVault did not become a widely adopted editing tool during our observation period. 
Instead, it substantially amplified the production capacity of a relatively small group of editors---mostly long-time editors---enabling them to generate a disproportionate share of newly added content of Korean Wikipedia. This content has received a very substantial fraction of pageviews.

\section{WikiVault Reshaped the Work of Producing and Maintaining Content}
Given that so few contributors were able to produce a strikingly higher volume of content, our investigation turned to how WikiVault changed \textit{the work} of contributors in producing content (RQ2). We found three key impacts of WikiVault: lowering barriers to contribution (\S\ref{sec:reduce_barrier}), shifting work from production toward review (\S\ref{sec:reorganizing_workflow}), and changing the editing dynamics of articles (\S\ref{sec:subseq_edit}). 

\subsection{Lowering Barriers Expanded the Work Editors Could Take On}
\label{sec:reduce_barrier}

Participants described WikiVault as reducing technical and linguistic barriers to contributing.
By automatically handling tasks like formatting wikitext, adapting templates, preserving references, and repairing links during translation, WikiVault reduced repetitive and tedious work. 
This not only reduced the burden on veteran editors, but also helped newcomers contribute without first mastering Wikipedia's technical conventions.
For instance, P6, a newcomer, reflected that without WikiVault, he might have simply \textit{``dropped out''} while trying to learn these conventions. 

WikiVault also reduced linguistic barriers at multiple stages of translation work. 
Rather than fully interpreting an English article before beginning to translate, editors could generate a Korean draft and develop their understanding as they revised it. 
As P3 described, \textit{``I skim the original, but first I click translate. Then the original and the translation appear together, so I read them side by side.''}
This also made unfamiliar domains more approachable to some participants. As P3 described, they could \textit{``work [through] it one piece at a time,''} noting that \textit{``you can really feel your knowledge building up as you translate''.}
Consequently, {WikiVault expanded the range of work that editors were willing and able to take on}, making previously daunting contributions more feasible.
P5, for example, recalled finally translating an article they had \textit{``intended to work on for 2 decades but had never felt able to tackle because of the anticipated effort.''}
P9 described how the hundreds of references in a rapidly evolving article made translating it manually \textit{``practically impossible,''} while WikiVault made the task manageable.

Yet, making more contributions actionable did not mean participants turned to WikiVault whenever they contributed.
Instead, the tool added a mode of contribution that editors could choose based on their bandwidth and emotional capacity. 
For P3, for example, manual translation was meaningful due to its challenge, preserving the \textit{``fun of think[ing] more deeply about the context [of a translation].''} Yet, WikiVault was convenient; if they were tired, they'd \textit{``just review [the WikiVault output] and get it posted,''} and \textit{``be happy with that.''} 
For P5, the choice also depended on the task. While they could readily edit existing articles manually, they noted that \textit{``[their] neurodivergence made starting a new article from scratch particularly difficult.''} 
The lower barrier could also make it easier for participants to act on work as it arose. For P9, translating articles often surfaced missing Korean Wikipedia pages through red links, which created an urge to fill them: \textit{``If you see all the red links, it bothers you and you want to quickly get rid of it.''} For P7, who focused on moderating content, WikiVault similarly made it \textit{``much more manageable''} to address existing articles that violated basic Wikipedia guidelines.

\subsection{Reorganizing the Relationship Between Production and Review}

\label{sec:reorganizing_workflow}
As participants created rough translation drafts with WikiVault, their work shifted toward revising and reviewing the output.
Yet, what this review entailed varied substantially.
For example, P6 triangulated between the English source article, Google Translate output, and WikiVault output, comparing multiple versions before deciding how to revise the article. But P4 often published WikiVault-output immediately and deferred closer inspection: \textit{``I usually just publish them right away. I publish them first, put the translation review on my personal queue, and plan to look at it later.''}
As P8 put it, \textit{``everyone has different standards for reviewing translated articles.''} 
P3 did not feel the need for extensive review since WikiVault's translation was difficult to distinguish from their own. 
For others, the tool's output was still awkward or ill-fitting language that required substantial manual intervention, producing text that \textit{``did not quite fit together''} (P5).
Review could also extend beyond the quality of the translation itself to verifying source material (P9) and further localizing content for Korean Wikipedia (P7).

\begin{table*}[t]
    \centering
    \caption{\textbf{Effects of WikiVault use on editing and content removal for existing and newly created articles.}
    Estimates are percentage-point differences from linear probability models
    with calendar-week fixed effects and standard errors clustered by matched pair.
    Brackets report 95\% confidence intervals.
    For existing articles (A), Week $-1$ is the omitted reference period, and joint-test $p$-values are from Wald tests that the coefficients for Weeks $-4$, $-3$, and $-2$ are jointly zero.
    The Week-0 non-bot editing outcome includes the WikiVault edit itself and therefore does not measure subsequent editing.
    For newly created articles (B), estimates compare WikiVault-created and matched non-WikiVault articles separately in each week after creation.}
    \Description{Table reporting estimated percentage-point effects of WikiVault use on the likelihood of non-bot editing and content removal for existing articles from four weeks before through four weeks after treatment. Joint tests do not reject zero pre-treatment effects for either outcome. In week 0, WikiVault use is associated with a 97.44 percentage-point increase in the likelihood of any non-bot edit and a 55.89 percentage-point increase in the likelihood of any bytes removed; the non-bot editing estimate includes the WikiVault edit itself. Positive effects on both outcomes decline substantially after week 0 but remain statistically significant through week 3, with week 4 estimates no longer statistically significant at the 0.05 level.}
    \label{tab:edit_removal_outcomes}
    \small

    \begin{minipage}[t]{0.48\textwidth}
    \centering
    \textbf{(A) Existing articles}\\[4pt]

    \begin{tabular}{lcc|cc}
        \toprule
        & \multicolumn{2}{c}{Any non-bot edit}
        & \multicolumn{2}{c}{Any bytes removed} \\
        \cmidrule(lr){2-3}
        \cmidrule(lr){4-5}
        & Estimate [95\% CI] & $p$
        & Estimate [95\% CI] & $p$ \\
        \midrule

        & \multicolumn{2}{c|}{Joint test: $p=.229$}
        & \multicolumn{2}{c}{Joint test: $p=.097$} \\
        \midrule

        Week $-4$
        & $-0.61$ [$-1.61$, $0.38$] & .229
        & $+0.00$ [$-0.72$, $0.73$] & .996 \\

        Week $-3$
        & $-0.90$ [$-1.85$, $0.04$] & .061
        & $-0.64$ [$-1.26$, $-0.01$] & .047 \\

        Week $-2$
        & $-0.86$ [$-1.77$, $0.05$] & .064
        & $-0.36$ [$-0.98$, $0.26$] & .251 \\

        Week $-1$
        & \multicolumn{2}{c|}{Reference}
        & \multicolumn{2}{c}{Reference} \\

        \midrule

        Week $0$
        & $+97.41$ [$96.65$, $98.17$] & $<.001$
        & $+55.87$ [$53.54$, $58.21$] & $<.001$ \\

        Week $1$
        & $+8.23$ [$6.64$, $9.81$] & $<.001$
        & $+5.40$ [$4.10$, $6.70$] & $<.001$ \\

        Week $2$
        & $+3.07$ [$1.78$, $4.36$] & $<.001$
        & $+1.70$ [$0.78$, $2.63$] & $<.001$ \\

        Week $3$
        & $+1.88$ [$0.67$, $3.09$] & .002
        & $+1.76$ [$0.83$, $2.69$] & $<.001$ \\

        Week $4$
        & $+1.07$ [$-0.05$, $2.20$] & .061
        & $+0.62$ [$-0.18$, $1.41$] & .128 \\

        \bottomrule
    \end{tabular}
    \end{minipage}
    \hfill
    %
    \begin{minipage}[t]{0.48\textwidth}
    \centering
    \textbf{(B) Newly created articles}\\[4pt]

    \begin{tabular}{lcc|cc}
        \toprule
        & \multicolumn{2}{c}{Any non-bot edit}
        & \multicolumn{2}{c}{Any bytes removed} \\
        \cmidrule(lr){2-3}
        \cmidrule(lr){4-5}
        & Estimate [95\% CI] & $p$
        & Estimate [95\% CI] & $p$ \\
        \midrule

        Week $0$
        & $+0.80$ [$-0.56$, $2.17$] & .250
        & $+13.65$ [$11.77$, $15.52$] & $<.001$ \\

        Week $1$
        & $-1.95$ [$-3.58$, $-0.32$] & .019
        & $+1.97$ [$0.88$, $3.06$] & $<.001$ \\

        Week $2$
        & $+0.71$ [$-0.10$, $1.52$] & .085
        & $+0.28$ [$-0.29$, $0.84$] & .344 \\

        Week $3$
        & $+0.76$ [$-0.02$, $1.53$] & .055
        & $+0.14$ [$-0.40$, $0.67$] & .613 \\

        Week $4$
        & $+0.57$ [$-0.11$, $1.25$] & .099
        & $+0.09$ [$-0.38$, $0.57$] & .706 \\

        Week $5$
        & $-0.55$ [$-1.20$, $0.10$] & .098
        & $-0.02$ [$-0.45$, $0.40$] & .916 \\

        Week $6$
        & $+0.60$ [$-0.05$, $1.24$] & .069
        & $+0.57$ [$0.13$, $1.02$] & .012 \\

        Week $7$
        & $+0.92$ [$0.28$, $1.55$] & .005
        & $+0.46$ [$0.01$, $0.90$] & .044 \\

        Week $8$
        & $+0.50$ [$-0.03$, $1.04$] & .065
        & $+0.46$ [$0.07$, $0.85$] & .020 \\

        Week $9$
        & $+0.76$ [$0.18$, $1.33$] & .010
        & $+0.64$ [$0.24$, $1.04$] & .002 \\

        Week $10$
        & $+0.23$ [$-0.30$, $0.76$] & .396
        & $+0.41$ [$0.04$, $0.79$] & .032 \\

        Week $11$
        & $+0.05$ [$-0.42$, $0.51$] & .848
        & $+0.14$ [$-0.19$, $0.47$] & .415 \\

        Week $12$
        & $+0.07$ [$-0.43$, $0.57$] & .789
        & $+0.02$ [$-0.33$, $0.38$] & .900 \\

        \bottomrule
    \end{tabular}
    \end{minipage}

    \vspace{4pt}
\end{table*}

Reviewing machine-assisted translations was not new to Korean Wikipedia, but WikiVault scaled this existing maintenance work.
Korean Wikipedia had long developed both norms and maintenance infrastructure around translated content, recommending revisions for unnatural translations%
\footnote{Translationese Phrases: https://ko.wikipedia.org/wiki/번역투\_문구}
and tracking articles whose translations had not yet been reviewed.%
\footnote{https://ko.wikipedia.org/wiki/분류:번역이\_검토되지\_않은\_문서}
Referencing these practices, P1 explained that \textit{``editors could identify obviously machine-translated articles and ask contributors to fix them,''} and that the community organized collective events to repair such articles when their volume had been manageable.
WikiVault, however, increased the production of content requiring review without a corresponding increase in review capacity.
As P9 described, \textit{``Machine-translation review issues have been an issue already, [...] we did expect this to happen, but the scale of it wasn't as expected.''} P4 pointed to how resulting backlog repeatedly raised concerns in Village Pump discussions, with P7 noting that \textit{``3,366 articles remain''} in the category for unreviewed translations.

As the review burden grew, the community's response shifted from absorbing the additional work to constraining how quickly WikiVault could produce.
Editors created a dedicated \textit{Unreviewed Translations (WikiVault)} subcategory\footnote{https://ko.wikipedia.org/wiki/분류:번역이\_검토되지\_않은\_문서\_(WikiVault)} to track WikiVault articles awaiting review, and formed a dedicated team to work through the backlog.
As it persisted, however, concerns shifted upstream toward how quickly WikiVault allowed new articles to be produced, leading P10 to fear \textit{``a more extreme proposal to ban AI altogether.''}
To them, WikiVault's benefits had become obvious, which led them to propose limiting how many translations each user could make per day and allowing administrators to restrict access for users who abused the tool.
Although these restrictions were technically straightforward to implement, the community treated them as a governance decision and deliberated on them in Village Pump discussions. Finally, on February 23, 2026, the community introduced a daily per-user limit on WikiVault calls.%
\footnote{https://ko.wikipedia.org/wiki/위키백과:사랑방/2026년\_제9주\#위키볼트\_이용\_제한\_기능\_도입}

\subsection{WikiVault Changed How Articles Were Subsequently Edited}
\label{sec:subseq_edit}


To examine how WikiVault affected subsequent editing, we compared WikiVault-edited articles with matched existing articles around the time of the WikiVault edit, and WikiVault-created articles with matched articles from the same creation period. 
We examine whether articles received subsequent non-bot edits and whether any content was removed, distinguishing between these two settings because WikiVault may affect existing and newly created articles differently. 

For \textit{existing} articles, WikiVault was followed by a sharp but short-lived increase in the likelihood of receiving subsequent edits (Table~\ref{tab:edit_removal_outcomes}A).
In the week following a WikiVault edit, treated articles were 8.2 percentage points more likely than matched controls to receive a non-bot edit (95\% CI [6.6, 9.8]).
This difference declined to 3.1 percentage points in week 2 and 1.9 percentage points in week 3, and was no longer statistically significant by week 4.
We observe a similar pattern for content removal.
Treated articles were 5.4 percentage points more likely to have content removed in the following week (95\% CI [4.1, 6.7]), with the difference declining to less than two percentage points in weeks 2 and 3 and becoming statistically indistinguishable from zero by week 4.
These suggest that the additional editing activity surrounding WikiVault edits was concentrated in the first few weeks after use.
Pre-treatment differences were small, at most one percentage point for either outcome (Table~\ref{tab:edit_removal_outcomes}A). 
The observed trajectories similarly show small differences between treated and matched articles before WikiVault use (Figure~\ref{fig:all_trajectories} E/G).
Joint tests did not reject that the three pre-treatment coefficients were jointly zero at the 5\% level for either editing ($p=.229$) or content removal ($p=.097$).

For \textit{newly created} articles, we did not observe a comparable increase in subsequent editing. 
WikiVault-created articles were 1.9 percentage points less likely than matched controls to receive a non-bot edit in the first week after creation (95\% CI [-3.6, -0.3]), while differences in later weeks were small and inconsistent.
WikiVault-created articles did, however, experience substantially more content removal immediately after creation.
During the creation week, they were 13.6 percentage points more likely than matched controls to have content removed (95\% CI [11.8, 15.5]).
This difference fell sharply to 2.0 percentage points in the following week (95\% CI [0.9, 3.1]) and remained smaller thereafter.

\section{WikiVault Reshaped What Knowledge Was Produced}
The changes in \textit{the work} producing and maintaining content also had consequences for \textit{what knowledge} was produced (RQ3). We first describe how WikiVault made it easier to carry English Wikipedia's editorial choices into Korean Wikipedia (\S\ref{sec:what_knowledge}), before examining how it unevenly changed different ways of producing knowledge and how editors and the community responded to these shifts (\S\ref{sec:balance_knowledge}). Last, we show how WikiVault changed the amount of content articles contained and the reader attention they received (\S\ref{sec:language-agnostic}).

\subsection{WikiVault Extended English Wikipedia's Editorial Influence}
\label{sec:what_knowledge}
A key observation across interviews was that {English Wikipedia and Korean Wikipedia varied in editorial norms}. 
P7 explained that descriptions acceptable under English Wikipedia Policies could be inappropriate; for example, evaluative descriptions like calling an athlete \textit{``one of the world's best,''} were, in their opinion, inconsistent with Korean Wikipedia's interpretation of Neutral Point of View policy even when sourced. 
Differences also concerned what information mattered and whose perspective an article represented. 
P1 described the English Wikipedia article on Malaysia as \textit{``well-written,''} while noting that it could differ both from how Malaysians saw their own country and from \textit{``what Korean Wikipedia editors think Korean readers need to know.’’}
These differences could also extend to meta-elements like article structure and naming conventions, as the two wikis had distinct local audiences. P9 described adding background when translating articles, noting that \textit{``English Wikipedia often assumes that readers already know the context and goes straight into the main content.''} 
P7 similarly described revising Central Asian names that English Wikipedia rendered through Russian-derived romanizations to reflect Uzbek or other local-language forms.

In that context, {WikiVault did not merely transfer content across languages but also the editorial choices embedded in how that content had been assembled and represented in English Wikipedia.} 
P7 described seeing more editors who regarded English Wikipedia standards and article style as the \textit{``correct''} ones after WikiVault's introduction:
\begin{quote}
    \textit{``I think Korean Wiki can benefit greatly from WikiVault when content brought over from English Wiki is adapted to the standards our community has established through consensus, and supplemented with information that editors research themselves. But as WikiVault has made it much easier to import English Wiki content, I have started seeing more editors who seem to treat English Wiki's infoboxes, article styles as 100\% correct---as if we should also write articles according to their standards.''}
\end{quote}
Editors already had practical reasons to trust English Wikipedia. P5, for example, pointed to the extensive sourcing in English Wikipedia, explaining that \textit{``even relatively niche articles like `Six the Musical' has around 120 references.''} P4 similarly described English Wikipedia editors as \textit{``well trained in evaluating sources''} and therefore, they did not generally question their references. 
Assuming those references were reliable was also practical because editors could not always verify them themselves. P3 checked references publicly available online but noted some materials requiring library access were often difficult to access from Korea, such that using WikiVault \textit{``starts from the assumption that the sources in the English Wikipedia article are reasonably sound. [...] Ultimately, you have to trust it.''} Participants suggested that over time, editors increasingly appeared to trust the editorial choices, article styles, and conventions of English Wikipedia as standards to follow in Korean Wikipedia. 

\subsection{WikiVault Shifted the Balance of Knowledge Production}

\label{sec:balance_knowledge}
WikiVault lowered barriers to contribution(\S\ref{sec:reduce_barrier}), but did so unevenly across different forms of knowledge production.
Pathways to produce knowledge through language editions other than English, external sources, and their own expertise remained substantially laborious.
P2, who regularly translated from English and Japanese Wikipedia, still had to \textit{translate Japanese articles often line by line,''} observing that \textit{about half [of the work] has gotten much better, but the other half is still about the same.''}
Other contributions required combining sources in ways WikiVault could not readily support. P5, for example, described translating an article available only in Chinese Wikipedia while consulting a German book they owned.
Finding and evaluating sources similarly remained largely with editors. P5 described this work as something \textit{``we humans have to do''} because it was \textit{``tacit and experiential,’’} while P1 described visiting museums to photograph materials for Korean topics when appropriate materials were unavailable online.

The limitations of relying on English Wikipedia were especially salient in  WikiVault's drafting feature, which allowed editors to produce initial drafts of articles when no English Wikipedia article existed. 
During a Korean food edit-a-thon, for example, P2 used WikiVault to draft an article on \textit{``gochujang-jjigae [고추장찌개]''}, a topic for which they were unsure how to structure an article because relevant materials were limited online.
The draft was a useful starting point, but 
the retrieved sources were blogs or columns that could not be directly cited on Wikipedia, leaving P2 to search for sources again: \textit{``I only used WikiVault to make the draft, and then basically had to start from scratch in finding the sources.''}
P4 similarly found that drafts provided \textit{only a sense and components of how an article might be developed,''} but often discarded them because the references were unreliable.
Consistent with these experiences, editors used drafting far less often than translation, as shown in \S\ref{subsec:adoption3}.
Drafting could therefore provide editors with an initial structure, but time-consuming manual work remained. 

As WikiVault changed what was easy to produce, some participants perceived community standards shifting toward contributions that the tool could not provide. 
P2 observed this shift in expectations for \textit{``Featured Articles''} and \textit{``Good Articles,''} two community-reviewed designations for high-quality articles, with the former representing a higher standard.\footnote{https://ko.wikipedia.org/wiki/위키백과:문서\_평가}
While translating and expanding high-quality English Wikipedia articles had been a common route to \textit{``Good Articles''}, P2 observed that \textit{``Featured Articles''} increasingly required \textit{``some distinctive feature that can only be found in Korean Wikipedia.''} This could mean \textit{``adding information about Korea or from Korean-language sources one by one,''} making it difficult, in P2’s view, \textit{``to be selected as a Featured Article just through WikiVault.''}

Editors and the broader community also responded by investing in knowledge that WikiVault could not readily provide. 
P1 traveled to Gongju and Buyeo to gather photographs and materials on local topics, focusing on contributions they were uniquely positioned to make. They also recalled concerns raised in the Village Pump about \textit{``the risk of Korean Wikipedia becoming a mirror of English Wikipedia,''} followed by a community-organized edit-a-thon focused on developing local content and localizing sources when appropriate. 
P9 similarly described initiatives by the Korean Wikimedia to develop Korean content---work for  \textit{``there is almost nothing we can use WikiVault for.''}
Participants saw these efforts as necessary to maintain language editions as distinct projects, with P9 emphasizing that \textit{``the project of building Korean Wikipedia and the project of building English Wikipedia are fundamentally different''}
while P1 pointed to similar efforts across Asian Wikimedia communities, like ESEAP,\footnote{
https://meta.wikimedia.org/wiki/ESEAP\_Hub} to strengthen knowledge grounded in local languages and communities in the age of AI.
As P1 summarized, \textit{``WikiVault itself is not the problem; what matters is how the community responds to it.''}

\begin{table*}[t]
    \centering
    \caption{\textbf{Effects of WikiVault use on pageviews and article length for existing and newly created articles.}
    Estimates are incidence rate ratios (IRRs) from Poisson models with
    calendar-week fixed effects and standard errors clustered by matched pair.
    Brackets report 95\% confidence intervals.
    For existing articles, the omitted reference period is the week immediately preceding WikiVault use ($t=-1$).
    For newly created articles, estimates compare WikiVault-created and matched non-WikiVault articles separately in each week after creation. Pre-treatment joint-test $p$-values are from Wald tests that the coefficients
    for weeks $-4$, $-3$, and $-2$ are jointly zero.}
    \Description{Table reporting differences in pageviews and article length for existing articles edited with WikiVault and newly created articles produced with WikiVault, compared with matched articles. For existing articles, pre-treatment estimates are close to an IRR of 1. Following WikiVault use, pageviews increase and remain elevated through week 4, while article length increases sharply in week 0 and remains substantially higher through week 4. WikiVault-created articles receive more pageviews shortly after creation, although this difference does not persist, and remain consistently about 65--68 percent longer than matched articles throughout the 12-week period.}
    \label{tab:pageviews_length}
    \small

    \begin{minipage}[t]{0.48\textwidth}
    \centering
    \textbf{(A) Existing articles}\\[4pt]

    \begin{tabular}{lcc|cc}
        \toprule
        & \multicolumn{2}{c}{Pageviews}
        & \multicolumn{2}{c}{Article length} \\
        \cmidrule(lr){2-3}
        \cmidrule(lr){4-5}
        & IRR [95\% CI] & $p$
        & IRR [95\% CI] & $p$ \\
        \midrule

        & \multicolumn{2}{c|}{Joint test: $p=.053$}
        & \multicolumn{2}{c}{Joint test: $p=.585$} \\
        \midrule

        Week $-4$
        & 1.022 [0.924, 1.130] & .676
        & 0.992 [0.934, 1.054] & .803 \\

        Week $-3$
        & 0.953 [0.882, 1.030] & .226
        & 0.990 [0.946, 1.036] & .666 \\

        Week $-2$
        & 0.970 [0.928, 1.014] & .182
        & 0.986 [0.965, 1.007] & .199 \\

        Week $-1$
        & \multicolumn{2}{c|}{Reference}
        & \multicolumn{2}{c}{Reference} \\

        \midrule

        Week $0$
        & 1.425 [1.296, 1.566] & $<.001$
        & 9.120 [8.407, 9.893] & $<.001$ \\

        Week $1$
        & 1.592 [1.040, 2.438] & .032
        & 4.422 [4.001, 4.887] & $<.001$ \\

        Week $2$
        & 1.604 [1.047, 2.458] & .030
        & 3.079 [2.732, 3.469] & $<.001$ \\

        Week $3$
        & 1.863 [1.015, 3.419] & .045
        & 2.778 [2.455, 3.144] & $<.001$ \\

        Week $4$
        & 1.994 [1.056, 3.766] & .033
        & 2.600 [2.312, 2.924] & $<.001$ \\

        \bottomrule
    \end{tabular}
    \end{minipage}
    \hfill
    \begin{minipage}[t]{0.48\textwidth}
    \centering
    \textbf{(B) Newly created articles}\\[4pt]

    \begin{tabular}{lcc|cc}
        \toprule
        & \multicolumn{2}{c}{Pageviews}
        & \multicolumn{2}{c}{Article length} \\
        \cmidrule(lr){2-3}
        \cmidrule(lr){4-5}
        & IRR [95\% CI] & $p$
        & IRR [95\% CI] & $p$ \\
        \midrule

        Week $0$
        & 1.284 [1.096, 1.503] & .002
        & 1.681 [1.573, 1.797] & $<.001$ \\

        Week $1$
        & 1.300 [1.015, 1.666] & .038
        & 1.662 [1.554, 1.778] & $<.001$ \\

        Week $2$
        & 1.296 [0.888, 1.891] & .179
        & 1.661 [1.553, 1.776] & $<.001$ \\

        Week $3$
        & 1.160 [0.823, 1.635] & .397
        & 1.658 [1.550, 1.773] & $<.001$ \\

        Week $4$
        & 0.963 [0.519, 1.786] & .905
        & 1.657 [1.549, 1.772] & $<.001$ \\

        Week $5$
        & 0.943 [0.402, 2.216] & .894
        & 1.655 [1.548, 1.770] & $<.001$ \\

        Week $6$
        & 0.941 [0.492, 1.798] & .853
        & 1.654 [1.546, 1.769] & $<.001$ \\

        Week $7$
        & 0.915 [0.378, 2.216] & .844
        & 1.653 [1.546, 1.768] & $<.001$ \\

        Week $8$
        & 0.705 [0.387, 1.283] & .252
        & 1.653 [1.545, 1.768] & $<.001$ \\

        Week $9$
        & 0.630 [0.373, 1.063] & .083
        & 1.652 [1.545, 1.767] & $<.001$ \\

        Week $10$
        & 0.685 [0.447, 1.051] & .083
        & 1.651 [1.544, 1.766] & $<.001$ \\

        Week $11$
        & 0.506 [0.280, 0.914] & .024
        & 1.650 [1.543, 1.764] & $<.001$ \\

        Week $12$
        & 0.513 [0.275, 0.957] & .036
        & 1.649 [1.542, 1.763] & $<.001$ \\

        \bottomrule
    \end{tabular}
    \end{minipage}
\end{table*}

\begin{figure*}[t]

    \includegraphics[width=0.875\linewidth, alt={Line chart showing weekly bytes added to Korean Wikipedia articles with and without a pre-existing English Wikipedia article from July 2024 to February 2026. Before WikiVault's deployment in May 2025, added bytes remain relatively stable for both groups. After deployment, added bytes to articles with a pre-existing English article increase sharply, with large peaks in mid-2025 and late 2025, while added bytes to articles without a pre-existing English article remain comparatively stable. A vertical line marks WikiVault's deployment.}]{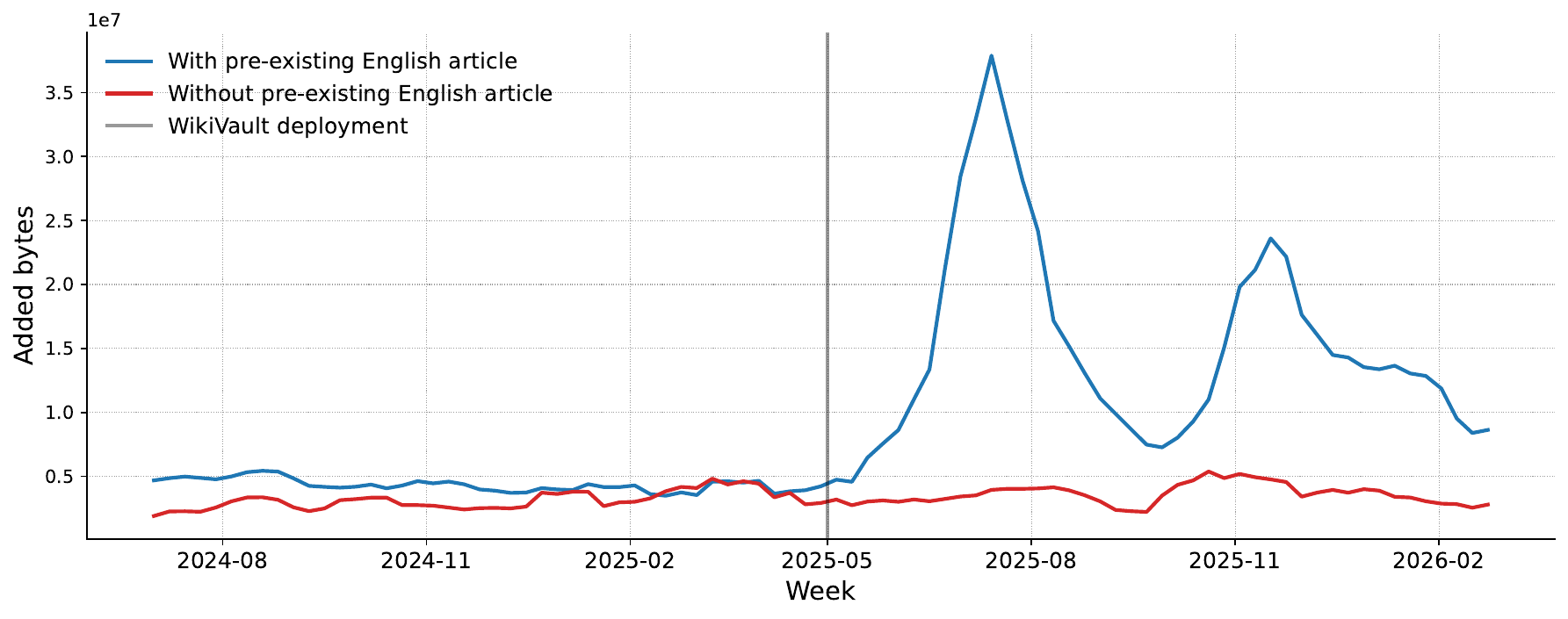}
            \caption{\textbf{Weekly added bytes to articles with and without a pre-existing English Wikipedia article.} The vertical line marks WikiVault's deployment.}
    \label{fig:coverage}
\end{figure*}

\subsection{WikiVault Produced Longer Articles and Received More Attention}
\label{sec:language-agnostic}
\label{sec:longer-articles}
Despite these uneven changes in what WikiVault made easier to produce (namely, making it easier to import content from English Wikipedia), we did not observe an aggregate decline in the production of articles unique to Korean Wikipedia. Figure~\ref{fig:coverage} shows that the production of Korean-Wikipedia-unique articles remained relatively stable over the observation period. Because WikiVault adoption remained concentrated among a small number of editors, however, these trends cannot tell us whether this balance would persist under broader adoption. 

We further assessed whether articles produced with WikiVault differed in form and attention using the same matched designs described previously (Table~\ref{tab:pageviews_length}). 
We found that WikiVault produced substantially longer articles, and these differences persisted beyond the initial contribution.
For existing articles, treated and matched articles followed similar trajectories in page length before the WikiVault edit (Table~\ref{tab:pageviews_length}A; Figure~\ref{fig:all_trajectories}C). 
During the intervention week, however, article length increased sharply relative to matched controls (IRR = 9.12, 95\% CI [8.41, 9.89], $p<.001$). 
This difference subsequently narrowed to 4.42 in week 1 and 3.08 in week 2, but remained substantial four weeks later (IRR = 2.60, 95\% CI [2.31, 2.92], $p<.001$).

For WikiVault-created articles, the difference in article length persisted with little change over time.
At creation, they were 68.1\% longer than matched articles created without WikiVault (IRR = 1.68, 95\% CI [1.57, 1.80], $p<.001$), and remained 64.9\% longer twelve weeks later (IRR = 1.65, 95\% CI [1.54, 1.76], $p<.001$). 
Together, this suggests that WikiVault contributions substantially expanded article content, much of which remained weeks or months later.

WikiVault-edited articles also received more reader attention, although this difference persisted only for existing articles.
For existing articles, treated and matched articles followed similar pageview trajectories before the WikiVault edit, with pre-treatment IRR close to 1. Following the WikiVault edit, pageviews increased and remained elevated throughout the following four weeks (Table~\ref{tab:pageviews_length}A; Figure ~\ref{fig:all_trajectories}A).
In the week of the edit, WikiVault-edited articles received 1.42 times as many pageviews as matched controls (95\% CI [1.30, 1.57], $p<.001$).
The difference remained statistically significant in each subsequent week, and by week 4, treated articles received 1.99 times as many pageviews as matched controls (95\% CI [1.06, 3.77], $p=.033$).

For newly created articles, higher pageviews were limited to the period shortly after creation (Table~\ref{tab:pageviews_length}B).
They received 28.4\% more pageviews than matched newly created articles in the week of creation (IRR = 1.28, 95\% CI [1.10, 1.50], $p=.002$) and 30.0\% more in the following week (IRR = 1.30, 95\% CI [1.01, 1.67], $p=.038$).
Estimates were no longer significantly higher after week 1 and were significantly lower than matched controls in weeks 11 and 12.
Thus, WikiVault-created articles remained substantially longer throughout the twelve weeks, but the initial difference in reader attention did not persist.

\section{Discussion}

Our findings illustrate how integrating LLMs into a digital knowledge commons can rebalance the interdependent forms of work that sustain it.
WikiVault substantially expanded the productive capacity of a small group of Korean Wikipedians, but this came at the cost of new demands for governing the resulting content.
At the same time, by making some forms of knowledge production easier than others, the tool shaped \textit{what} knowledge was produced and whose editorial choices traveled across language editions.
In this discussion, we draw on these dynamics to consider what LLMs mean for the sustainability of knowledge commons more broadly, and how communities can navigate the adoption of emerging (AI) technologies whose consequences are difficult to anticipate.

\subsection{Sustaining a Larger Commons with Fewer Contributors?}

Sustaining Wikipedia has long depended on maintaining a sufficiently large and active contributor base.
Our findings show that WikiVault expanded the commons not by bringing substantially more people into the community, but by enabling a small number of experienced contributors to produce much more (\S\ref{subsec:adoption3}).
WikiVault therefore raises a possibility that would have seemed implausible under earlier models of peer production: that a much smaller group of contributors might produce a body of knowledge that previously required many more.  At first blush, this observation promises an alternative strategy to sustain digital commons over time. Participation in commons-based peer production projects like Wikipedia shows slower growth over time \cite{halfaker_rise_2013,teblunthuis_revisiting_2018} and continues to wrestle with questions of newcomer recruitment and retention \cite{morgan2018evaluating}. Prior research attributes this slowing down at least in part to rules and processes that become entrenched and consequently produce barriers to entry \cite{halfaker_rise_2013,hill2020wikipedia,halfaker_dont_2011}. 

However, producing a larger commons is not necessarily the same as becoming a sustainable one, as expanding a knowledge commons also expands what the community must govern. 
 As WikiVault increased the production of new content in Korean Wikipedia, our findings show how contributors did more subsequent editing (\S\ref{sec:subseq_edit}) and how editors collectively developed new routines for reviewing and managing that content (\S\ref{sec:reorganizing_workflow}). In this sense, WikiVault relaxed a constraint on \textit{producing} the commons while making constraints on \textit{governing} it increasingly salient. This governance challenge extended beyond verifying the quality of individual contributions to questions about what kind of knowledge commons Korean Wikipedia should be. 
WikiVault's reliance on English Wikipedia prompted critical reflection on Korean Wikipedia's identity and place as a distinct project, echoing prior empirical work showing how Wikipedia communities reflect diverse knowledge representations \cite{hecht_tower_2010}. 
In this sense, governing WikiVault's output also became a process of defining --- and defending --- the boundaries of Korean Wikipedia as a distinct knowledge project. Translations from English Wikipedia carried not only content but also judgments about what was relevant, whose perspectives were represented, what context readers needed, and how subjects should be described (\S\ref{sec:what_knowledge}). Participants responded by adapting these choices to local norms, investing in knowledge that WikiVault could not readily produce, and developing constraints on how the tool could be used.

We note two important implications of the shift toward governance and review. First, the transfer of content enabled by WikiVault created a new starting point for peer production work for certain types of content (i.e., what's available on English Wikipedia) and not others. However, participation in a knowledge commons like Wikipedia remains voluntary, wherein contributors choose what work to take on and how much of themselves to invest in it \cite{benkler_coases_2002}.
WikiVault did not always align with contributors' topical interests and also required a willingness to engage in multilingual content review. Thus, whether the type of contribution the tool enabled was more or less enjoyable or compelling to the community remains an open question. 
Second, although our findings suggest WikiVault lowered barriers to contribution, governing that content required deeper knowledge of and commitment to Korean Wikipedia.  Most WikiVault users were experienced, long-time contributors who understood the wiki's specific norms and policies.
This raises a potential paradox: WikiVault relies on expertise that contributors developed through forms of participation that the tool itself may increasingly displace. As \citet{madison2024knowledge} observes, one important product of Wikipedia is the Wikipedia community itself; indeed, knowledge commons ``is’’ governance. And in a world where tools like WikiVault become a primary pathway to contribution, it is unclear how communities will produce the experienced contributors they need. Thus, whether fewer contributors can sustain a larger commons depends not only on what contributors \textit{can} and \textit{want} to produce, but also on whether the commons can keep producing the people capable of governing it.


\subsection{Rolling Out Emerging (AI) Technologies} 

Communities face difficult decisions about whether and how to adopt emerging technologies whose consequences cannot be fully known in advance~\cite{genus2018collingridge}.
This problem is particularly acute for generative AI, whose adoption has generated substantial controversy across Wikipedia communities.
In June 2025, for example, the Wikimedia Foundation paused a trial of AI-generated article summaries within days after strong editor opposition.\footnote{\url{https://en.wikipedia.org/wiki/Wikipedia:Village\_pump\_(technical)/Archive\_221\#Simple\_summaries:\_editor\_survey\_and\_2-week\_mobile\_study}}
Korean Wikipedia was no exception. 
We found skepticism toward AI both preceded and persisted after WikiVault's release, shaping decisions to avoid broad debate during development~(\S\ref{subsec:adoption2}) and later raising concerns that controversy over its misuse could escalate into a blanket restriction on AI~(\S\ref{sec:reorganizing_workflow}).
The tension and sense of potential controversy creates uncertainty for communities who are trying to navigate whether and how to integrate emerging technologies in their spaces.

Longstanding HCI work on technology appropriation and domestication shows that technologies integrate into social settings through mutual adaptation among artifacts, practices, and meanings~\cite{dourish_appropriation_2003, berker2005domestication}. That is, the consequences of an emerging technology may become apparent only through actual use. 
Thus, we present WikiVault as a case study that shows what can become knowable through such use. 
In Korean Wikipedia, community members themselves introduced a new technological capacity into their commons and learned from and reflected on what unfolded around it: from what the technology could do, to the consequences of scaling those capabilities, and, ultimately, to what it meant for Korean Wikipedia itself.

Yet introducing a technology into practice does not by itself ensure collective learning and reflection from that experience. We suggest that a few important conditions enabled the community's reflection and iteration. First, WikiVault was a \textit{community}-developed tool, which not only established legitimacy within the community but also allowed it to target core contribution tasks like translation that are not necessarily central to all language editions of Wikipedia. 
Second, WikiVault's developers intentionally limited the tool's availability to prevent a huge influx of users generating WikiVault-produced content at scale. Initially constraining the roll-out allowed the tool to be introduced as an experiment that would not suddenly overwhelm the community. Whether WikiVault was ultimately beneficial for Korean Wikipedia is a separate and more complicated question --- and one that likely can only be judged over time --- but its actual deployment became {productive as a way of learning} what its introduction could mean for the community. 
Third, WikiVault use was often legible to the community, as contributions made through it were automatically tagged. Although contributors could technically bypass the tag, WikiVault use thus largely remained visible through this shared infrastructure. Contributors could track unreviewed WikiVault translations through a dedicated category page, observe growing review burdens, and discuss when and how its use should be constrained. This traceability gave the community something concrete to collectively reason about and organize around. Indeed, this same traceability made our own quantitative analysis possible, allowing us to identify WikiVault-assisted edits and examine their consequences at scale. 

Together, these conditions --- legitimacy, limitations, and legibility --- created space for reflexivity about the community itself, and the role and place of an LLM-based technology within it. Our findings highlighted how, through assessing and reflecting on WikiVault, contributors reconsidered what should distinguish Korean Wikipedia as its own knowledge project~(\S\ref{sec:balance_knowledge}). 
The case of WikiVault thus suggests that other communities seeking to integrate novel AI features might follow a similar strategy (legitimacy, limitations, legibility) to assess and iterate development and norm-setting. However, we also recognize that this is not a costless process. In developing and deploying WikiVault, the local community was deeply invested in experimenting with the tool and even provided financial support. As an iterative process, experimenting with and reflecting on WikiVault required contributors willing to observe consequences, reflect on what they mean, and translate those lessons into new norms and practices. Whether other contexts have similar values and incentive structures is likely to impact how similar technologies are rolled out and received there.

\subsection{Limitations and Future Work}
Our study has various limitations that suggest directions for future work.
First, WikiVault is a single case situated within 
Korean Wikipedia, where contributor scarcity and translation from English Wikipedia were particularly salient.
Future work could examine how such tools unfold in communities with different needs, norms, and governance structures.
Second, our findings primarily reflect AI-assisted translation by experienced contributors, limiting what we can conclude about LLM-assisted original drafting or use by newcomers at scale. Our interviews also primarily capture people who directly engaged with WikiVault, leaving out the perspectives of editors who avoided the tool, which are less represented.
Finally, our quantitative analyses do not directly assess the epistemic consequences of WikiVault-produced content.
Future work could examine how interlinked articles across language editions converge or diverge over time in their content and sources, and whether LLM-assisted translation closes existing knowledge gaps or reproduces asymmetries between language editions.

\section{Conclusion}

This paper examined what happened when the Korean Wikipedia community developed and deployed an LLM-based tool.
WikiVault enabled a small group of contributors to produce substantially more content by translating from English Wikipedia.
Yet this increase in production did not eliminate the human work required to sustain the resulting knowledge.
Instead, it shifted that work toward reviewing, verifying, localizing, maintaining, and governing what the tool made easier to produce. 
The tool also changed how easily different kinds of knowledge could enter Korean Wikipedia, prompting contributors to reconsider what should remain 
distinctive about their own knowledge project.
More broadly, WikiVault shows that the governance of emerging technologies need not be settled before adoption.
Contributors discovered the tool's consequences through use, made those consequences visible to one another, and revised their practices and governance in response.
Because WikiVault's use was legitimate, visible, and revisable, its effects became objects of collective scrutiny rather than hidden individual experiments.


\begin{acks}
We thank the participants who generously shared their time, experiences, and perspectives with us. We are also grateful to the Korean Wikipedia community and Wikimedia Korea for their openness and support throughout this research. Finally, we thank Pablo Aragón for his continued feedback and guidance, which were invaluable in shaping this work.
\end{acks}
\bibliographystyle{ACM-Reference-Format}
\bibliography{sample-base}

@inproceedings{vincent2018examining,
  title={Examining Wikipedia with a broader lens: Quantifying the value of Wikipedia's relationships with other large-scale online communities},
  author={Vincent, Nicholas and Johnson, Isaac and Hecht, Brent},
  booktitle={Proceedings of the 2018 CHI Conference on Human Factors in Computing Systems},
  pages={1--13},
  year={2018}
}

@article{walters2023fabrication,
  title={Fabrication and errors in the bibliographic citations generated by ChatGPT},
  author={Walters, William H and Wilder, Esther Isabelle},
  journal={Scientific reports},
  volume={13},
  number={1},
  pages={14045},
  year={2023},
  publisher={Nature Publishing Group UK London}
}

@article{jakesch2023human,
  title={Human heuristics for AI-generated language are flawed},
  author={Jakesch, Maurice and Hancock, Jeffrey T and Naaman, Mor},
  journal={Proceedings of the National Academy of Sciences},
  volume={120},
  number={11},
  pages={e2208839120},
  year={2023},
  publisher={National Academy of Sciences}
}

@inproceedings{clark2021all,
  title={All that’s ‘human’is not gold: Evaluating human evaluation of generated text},
  author={Clark, Elizabeth and August, Tal and Serrano, Sofia and Haduong, Nikita and Gururangan, Suchin and Smith, Noah A},
  booktitle={Proceedings of the 59th Annual Meeting of the Association for Computational Linguistics and the 11th International Joint Conference on Natural Language Processing (Volume 1: Long Papers)},
  pages={7282--7296},
  year={2021}
}

@misc{miller_new_2025,
	title = {New {User} {Trends} on {Wikipedia}},
	url = {https://diff.wikimedia.org/2025/10/17/new-user-trends-on-wikipedia/},
	language = {en-US},
	urldate = {2026-09-07},
	journal = {Diff},
	author = {Miller, Marshall},
	month = oct,
	year = {2025},
}

@article{cohen_wikipedia_2021,
	title = {Wikipedia {Is} {Finally} {Asking} {Big} {Tech} to {Pay} {Up}},
	issn = {1059-1028},
	url = {https://www.wired.com/story/wikipedia-finally-asking-big-tech-to-pay-up/},
	language = {en-US},
	urldate = {2026-09-07},
	journal = {Wired},
	author = {Cohen, Noam},
	month = mar,
	year = {2021},
	note = {Section: tags},
}

@article{vincent2021deeper,
  title={A deeper investigation of the importance of Wikipedia links to search engine results},
  author={Vincent, Nicholas and Hecht, Brent},
  journal={Proceedings of the ACM on Human-Computer Interaction},
  volume={5},
  number={CSCW1},
  pages={1--15},
  year={2021},
  publisher={ACM New York, NY, USA}
}

@article{buccinca2021trust,
  title={To trust or to think: cognitive forcing functions can reduce overreliance on AI in AI-assisted decision-making},
  author={Bu{\c{c}}inca, Zana and Malaya, Maja Barbara and Gajos, Krzysztof Z},
  journal={Proceedings of the ACM on Human-computer Interaction},
  volume={5},
  number={CSCW1},
  pages={1--21},
  year={2021},
  publisher={ACM New York, NY, USA}
}

@misc{wiki_de_kiverbot,
	title = {{Wikipedia:Meinungsbilder/KI-Verbot}},
	howpublished = {German-language Wikipedia},
	url = {https://de.wikipedia.org/wiki/Wikipedia:Meinungsbilder/KI-Verbot},
	author = {{Wikipedia contributors}},
	year = {2026},
	note = {Accessed: 2026-06-30},
}

@article{jackson201411,
  title={11 rethinking repair},
  author={Jackson, Steven J},
  journal={Media technologies: Essays on communication, materiality, and society},
  pages={221--39},
  year={2014},
  publisher={MIT press Cambridge, MA, USA}
}

@article{dourish_appropriation_2003,
	title = {The {Appropriation} of {Interactive} {Technologies}: {Some} {Lessons} from {Placeless} {Documents}},
	volume = {12},
	number = {4},
	journal = {Computer Supported Cooperative Work (CSCW)},
	author = {Dourish, Paul},
	year = {2003},
	pages = {465--490},
}

@inproceedings{laxstrom2015content,
  title={Content Translation: Computer-assisted translation tool for Wikipedia articles},
  author={Laxstr{\"o}m, Niklas and Giner, Pau and Thottingal, Santhosh},
  booktitle={Proceedings of the 18th Annual Conference of the European Association for Machine Translation},
  year={2015}
}

@article{tsvetkova2017even,
  title={Even good bots fight: The case of Wikipedia},
  author={Tsvetkova, Milena and Garc{\'\i}a-Gavilanes, Ruth and Floridi, Luciano and Yasseri, Taha},
  journal={PloS one},
  volume={12},
  number={2},
  pages={e0171774},
  year={2017},
  publisher={Public Library of Science San Francisco, CA USA}
}

@article{halfaker2012bots,
  title={Bots and cyborgs: Wikipedia's immune system},
  author={Halfaker, Aaron and Riedl, John},
  journal={Computer},
  volume={45},
  number={03},
  pages={79--82},
  year={2012},
  publisher={IEEE Computer Society}
}

@inproceedings{hecht_tower_2010,
	title = {The {Tower} of {Babel} {Meets} {Web} 2.0: {User}-{Generated} {Content} and {Its} {Applications} in a {Multilingual} {Context}},
	booktitle = {Proceedings of the {SIGCHI} {Conference} on {Human} {Factors} in {Computing} {Systems}},
	publisher = {ACM},
	author = {Hecht, Brent and Gergle, Darren},
	year = {2010},
	pages = {291--300},
}

@inproceedings{bao_omnipedia_2012,
	title = {Omnipedia: bridging the {Wikipedia} language gap},
	booktitle = {Proceedings of the {SIGCHI} {Conference} on {Human} {Factors} in {Computing} {Systems}},
	publisher = {ACM},
	author = {Bao, Patti and Hecht, Brent and Carton, Samuel and Quaderi, Mahmood and Horn, Michael and Gergle, Darren},
	year = {2012},
	pages = {1075--1084},
}

@inproceedings{samir_locating_2024,
  title={Locating information gaps and narrative inconsistencies across languages: a case study of LGBT people portrayals on Wikipedia},
  author={Samir, Farhan and Park, Chan Young and Field, Anjalie and Shwartz, Vered and Tsvetkov, Yulia},
  booktitle={Proceedings of the 2024 Conference on Empirical Methods in Natural Language Processing},
  pages={6747--6762},
  year={2024}
}

@inproceedings{wulczyn_growing_2016,
	title = {Growing {Wikipedia} {Across} {Languages} via {Recommendation}},
	booktitle = {Proceedings of the 25th {International} {Conference} on {World} {Wide} {Web}},
	author = {Wulczyn, Ellery and West, Robert and Zia, Leila and Leskovec, Jure},
	year = {2016},
	pages = {975--985},
}

@article{schmidt1992taking,
  title={Taking CSCW seriously: Supporting articulation work},
  author={Schmidt, Kjeld and Bannon, Liam},
  journal={Computer supported cooperative work (CSCW)},
  volume={1},
  number={1},
  pages={7--40},
  year={1992},
  publisher={Springer}
}

@article{star_layers_1999,
	title = {Layers of {Silence}, {Arenas} of {Voice}: {The} {Ecology} of {Visible} and {Invisible} {Work}},
	volume = {8},
	number = {1-2},
	journal = {Computer Supported Cooperative Work (CSCW)},
	author = {Star, Susan Leigh and Strauss, Anselm},
	year = {1999},
	pages = {9--30},
}

@article{star_steps_1996,
	title = {Steps {Toward} an {Ecology} of {Infrastructure}: {Design} and {Access} for {Large} {Information} {Spaces}},
	volume = {7},
	number = {1},
	journal = {Information Systems Research},
	author = {Star, Susan Leigh and Ruhleder, Karen},
	year = {1996},
	pages = {111--134},
}

@article{benkler_coases_2002,
	title = {Coase's {Penguin}, or, {Linux} and ``{The} {Nature} of the {Firm}''},
	volume = {112},
	issn = {0044-0094},
	url = {https://www.jstor.org/stable/1562247},
	doi = {10.2307/1562247},
	number = {3},
	urldate = {2026-06-16},
	journal = {The Yale Law Journal},
	publisher = {The Yale Law Journal Company, Inc.},
	author = {Benkler, Yochai},
	year = {2002},
	pages = {369--446},
}

@book{benkler_wealth_2006,
	title = {The {Wealth} of {Networks}: {How} {Social} {Production} {Transforms} {Markets} and {Freedom}},
	isbn = {978-0-300-11056-2},
	shorttitle = {The {Wealth} of {Networks}},
	url = {https://www.jstor.org/stable/j.ctt1njknw},
	urldate = {2026-06-05},
	publisher = {Yale University Press},
	author = {Benkler, Yochai},
	year = {2006},
}

@book{ostrom1990governing,
  title={Governing the commons: The evolution of institutions for collective action},
  author={Ostrom, Elinor},
  year={1990},
  publisher={Cambridge University Press}
}

@book{hess_understanding_2007,
  title={Understanding Knowledge as a Commons: From Theory to Practice},
  editor={Hess, Charlotte and Ostrom, Elinor},
  year={2007},
  publisher={MIT Press}
}

@book{frischmann_governing_2014,
	title = {Governing {Knowledge} {Commons}},
	isbn = {978-0-19-022582-7},
	language = {en},
	publisher = {Oxford University Press},
	author = {Frischmann, Brett M. and Madison, Michael J. and Strandburg, Katherine Jo},
	year = {2014},
}

@book{kraut_building_2012,
	title = {Building {Successful} {Online} {Communities}: {Evidence}-{Based} {Social} {Design}},
	isbn = {978-0-262-29739-4},
	shorttitle = {Building {Successful} {Online} {Communities}},
	language = {en},
	publisher = {MIT Press},
	author = {Kraut, Robert E. and Resnick, Paul},
	month = mar,
	year = {2012},
}

@article{ren_how_2023,
	title = {How {Did} {They} {Build} the {Free} {Encyclopedia}? {A} {Literature} {Review} of {Collaboration} and {Coordination} among {Wikipedia} {Editors}},
	volume = {31},
	issn = {1073-0516},
	shorttitle = {How {Did} {They} {Build} the {Free} {Encyclopedia}?},
	url = {https://dl.acm.org/doi/10.1145/3617369},
	doi = {10.1145/3617369},
	number = {1},
	urldate = {2026-06-16},
	journal = {ACM Transactions on Computer-Human Interaction},
	author = {Ren, Yuqing and Zhang, Haifeng and Kraut, Robert E.},
	month = nov,
	year = {2023},
	pages = {7:1--7:48},
}

@article{kittur2007power,
  title={Power of the few vs. wisdom of the crowd: Wikipedia and the rise of the bourgeoisie},
  author={Kittur, A and Chi, E and Pendleton, BA and Suh, B and Mytkowicz, T},
  journal={World Wide Web},
  volume={1},
  pages={19},
  year={2007}
}

@inproceedings{kittur_he_2007,
	address = {New York, NY, USA},
	series = {{CHI} '07},
	title = {He says, she says: conflict and coordination in {Wikipedia}},
	isbn = {978-1-59593-593-9},
	url = {https://doi.org/10.1145/1240624.1240698},
	doi = {10.1145/1240624.1240698},
	booktitle = {Proceedings of the {SIGCHI} {Conference} on {Human} {Factors} in {Computing} {Systems}},
	publisher = {Association for Computing Machinery},
	author = {Kittur, Aniket and Suh, Bongwon and Pendleton, Bryan A. and Chi, Ed H.},
	year = {2007},
	pages = {453--462},
}

@inproceedings{geiger_work_2010,
	address = {New York, NY, USA},
	series = {{CSCW} '10},
	title = {The work of sustaining order in {Wikipedia}: the banning of a vandal},
	isbn = {978-1-60558-795-0},
	url = {https://doi.org/10.1145/1718918.1718941},
	doi = {10.1145/1718918.1718941},
	booktitle = {Proceedings of the 2010 {ACM} {Conference} on {Computer} {Supported} {Cooperative} {Work}},
	publisher = {Association for Computing Machinery},
	author = {Geiger, R. Stuart and Ribes, David},
	year = {2010},
	pages = {117--126},
}

@inproceedings{geiger_when_2013,
	address = {New York, NY, USA},
	series = {{WikiSym} '13},
	title = {When the levee breaks: without bots, what happens to {Wikipedia}'s quality control processes?},
	isbn = {978-1-4503-1852-5},
	shorttitle = {When the levee breaks},
	url = {https://dl.acm.org/doi/10.1145/2491055.2491061},
	doi = {10.1145/2491055.2491061},
	urldate = {2026-06-16},
	booktitle = {Proceedings of the 9th {International} {Symposium} on {Open} {Collaboration}},
	publisher = {Association for Computing Machinery},
	author = {Geiger, R. Stuart and Halfaker, Aaron},
	month = aug,
	year = {2013},
	pages = {1--6},
}

@inproceedings{halfaker_dont_2011,
	address = {New York, NY, USA},
	series = {{WikiSym} '11},
	title = {Don't bite the newbies: how reverts affect the quantity and quality of {Wikipedia} work},
	isbn = {978-1-4503-0909-7},
	shorttitle = {Don't bite the newbies},
	url = {https://dl.acm.org/doi/10.1145/2038558.2038585},
	doi = {10.1145/2038558.2038585},
	urldate = {2026-06-16},
	booktitle = {Proceedings of the 7th {International} {Symposium} on {Wikis} and {Open} {Collaboration}},
	publisher = {Association for Computing Machinery},
	author = {Halfaker, Aaron and Kittur, Aniket and Riedl, John},
	month = oct,
	year = {2011},
	pages = {163--172},
}

@inproceedings{forte_scaling_2008,
	title = {Scaling {Consensus}: {Increasing} {Decentralization} in {Wikipedia} {Governance}},
	issn = {1530-1605},
	shorttitle = {Scaling {Consensus}},
	url = {https://ieeexplore.ieee.org/abstract/document/4438860},
	doi = {10.1109/HICSS.2008.383},
	urldate = {2026-06-16},
	booktitle = {Proceedings of the 41st {Annual} {Hawaii} {International} {Conference} on {System} {Sciences} ({HICSS} 2008)},
	author = {Forte, Andrea and Bruckman, Amy},
	month = jan,
	year = {2008},
	pages = {157--157},
}

@inproceedings{swarts_collaborative_2009,
	title = {The Collaborative Construction of ``Fact'' on {Wikipedia}},
	pages = {281--288},
	booktitle = {Proceedings of the 27th ACM International Conference on Design of Communication},
	author = {Swarts, Jason},
	year = {2009},
}

@article{halfaker_rise_2013,
	title = {The {Rise} and {Decline} of an {Open} {Collaboration} {System}: {How} {Wikipedia}’s {Reaction} to {Popularity} {Is} {Causing} {Its} {Decline}},
	volume = {57},
	issn = {0002-7642},
	shorttitle = {The {Rise} and {Decline} of an {Open} {Collaboration} {System}},
	url = {https://doi.org/10.1177/0002764212469365},
	doi = {10.1177/0002764212469365},
	number = {5},
	urldate = {2026-06-05},
	journal = {American Behavioral Scientist},
	publisher = {SAGE Publications Inc},
	author = {Halfaker, Aaron and Geiger, R. Stuart and Morgan, Jonathan T. and Riedl, John},
	month = may,
	year = {2013},
	pages = {664--688},
}

@inproceedings{teblunthuis_revisiting_2018,
	address = {New York, NY, USA},
	series = {{CHI} '18},
	title = {Revisiting ``{The} {Rise} and {Decline}'' in a {Population} of {Peer} {Production} {Projects}},
	isbn = {978-1-4503-5620-6},
	url = {https://dl.acm.org/doi/10.1145/3173574.3173929},
	doi = {10.1145/3173574.3173929},
	urldate = {2026-06-16},
	booktitle = {Proceedings of the 2018 {CHI} {Conference} on {Human} {Factors} in {Computing} {Systems}},
	publisher = {Association for Computing Machinery},
	author = {TeBlunthuis, Nathan and Shaw, Aaron and Hill, Benjamin Mako},
	month = apr,
	year = {2018},
	pages = {1--7},
}

@article{zheng_roles_2019,
	address = {New York, NY, USA},
	title = {The {Roles} {Bots} {Play} in {Wikipedia}},
	volume = {3},
	url = {https://doi.org/10.1145/3359317},
	doi = {10.1145/3359317},
	number = {CSCW},
	journal = {Proceedings of the ACM on Human-Computer Interaction},
	publisher = {Association for Computing Machinery},
	author = {Zheng, Lei (Nico) and Albano, Christopher M. and Vora, Neev M. and Mai, Feng and Nickerson, Jeffrey V.},
	month = nov,
	year = {2019},
}

@article{halfaker_ores_2020,
	address = {New York, NY, USA},
	title = {{ORES}: {Lowering} {Barriers} with {Participatory} {Machine} {Learning} in {Wikipedia}},
	volume = {4},
	url = {https://doi.org/10.1145/3415219},
	doi = {10.1145/3415219},
	number = {CSCW2},
	journal = {Proceedings of the ACM on Human-Computer Interaction},
	publisher = {Association for Computing Machinery},
	author = {Halfaker, Aaron and Geiger, R. Stuart},
	month = oct,
	year = {2020},
}

@inproceedings{adler_content-driven_2007,
	address = {New York, NY, USA},
	series = {{WWW} '07},
	title = {A content-driven reputation system for the {Wikipedia}},
	isbn = {978-1-59593-654-7},
	url = {https://dl.acm.org/doi/10.1145/1242572.1242608},
	doi = {10.1145/1242572.1242608},
	urldate = {2026-07-04},
	booktitle = {Proceedings of the 16th international conference on {World} {Wide} {Web}},
	publisher = {Association for Computing Machinery},
	author = {Adler, B. Thomas and de Alfaro, Luca},
	month = may,
	year = {2007},
	pages = {261--270},
}

@inproceedings{blumenstock_size_2008,
	address = {New York, NY, USA},
	series = {{WWW} '08},
	title = {Size matters: word count as a measure of quality on {Wikipedia}},
	isbn = {978-1-60558-085-2},
	shorttitle = {Size matters},
	url = {https://dl.acm.org/doi/10.1145/1367497.1367673},
	doi = {10.1145/1367497.1367673},
	urldate = {2026-07-04},
	booktitle = {Proceedings of the 17th international conference on {World} {Wide} {Web}},
	publisher = {Association for Computing Machinery},
	author = {Blumenstock, Joshua E.},
	month = apr,
	year = {2008},
	pages = {1095--1096},
}

@inproceedings{warncke-wang_tell_2013,
	address = {New York, NY, USA},
	series = {{WikiSym} '13},
	title = {Tell me more: an actionable quality model for {Wikipedia}},
	isbn = {978-1-4503-1852-5},
	shorttitle = {Tell me more},
	url = {https://dl.acm.org/doi/10.1145/2491055.2491063},
	doi = {10.1145/2491055.2491063},
	urldate = {2026-07-04},
	booktitle = {Proceedings of the 9th {International} {Symposium} on {Open} {Collaboration}},
	publisher = {Association for Computing Machinery},
	author = {Warncke-Wang, Morten and Cosley, Dan and Riedl, John},
	month = aug,
	year = {2013},
	pages = {1--10},
}

@article{moas_automatic_2023,
	title = {Automatic Quality Assessment of {Wikipedia} Articles---A Systematic Literature Review},
	volume = {56},
	number = {4},
	pages = {1--37},
	journal = {ACM Computing Surveys},
	author = {Mo\'{a}s, Pedro Miguel and Lopes, Carla Teixeira},
	year = {2023},
}

@article{jhaver_human-machine_2019,
	title = {Human-{Machine} {Collaboration} for {Content} {Regulation}: {The} {Case} of {Reddit} {Automoderator}},
	volume = {26},
	issn = {1073-0516},
	shorttitle = {Human-{Machine} {Collaboration} for {Content} {Regulation}},
	url = {https://dl.acm.org/doi/10.1145/3338243},
	doi = {10.1145/3338243},
	number = {5},
	urldate = {2026-06-05},
	journal = {ACM Transactions on Computer-Human Interaction (TOCHI)},
	author = {Jhaver, Shagun and Birman, Iris and Gilbert, Eric and Bruckman, Amy},
	month = jul,
	year = {2019},
	pages = {31:1--31:35},
}

@article{schluger_proactive_2022,
	address = {New York, NY, USA},
	title = {Proactive {Moderation} of {Online} {Discussions}: {Existing} {Practices} and the {Potential} for {Algorithmic} {Support}},
	volume = {6},
	url = {https://doi.org/10.1145/3555095},
	doi = {10.1145/3555095},
	number = {CSCW2},
	journal = {Proceedings of the ACM on Human-Computer Interaction},
	publisher = {Association for Computing Machinery},
	author = {Schluger, Charlotte and Chang, Jonathan P. and Danescu-Niculescu-Mizil, Cristian and Levy, Karen},
	month = nov,
	year = {2022},
}

@inproceedings{smith_keeping_2020,
	address = {New York, NY, USA},
	series = {{CHI} '20},
	title = {Keeping {Community} in the {Loop}: {Understanding} {Wikipedia} {Stakeholder} {Values} for {Machine} {Learning}-{Based} {Systems}},
	isbn = {978-1-4503-6708-0},
	url = {https://doi.org/10.1145/3313831.3376783},
	doi = {10.1145/3313831.3376783},
	booktitle = {Proceedings of the 2020 {CHI} {Conference} on {Human} {Factors} in {Computing} {Systems}},
	publisher = {Association for Computing Machinery},
	author = {Smith, C. Estelle and Yu, Bowen and Srivastava, Anjali and Halfaker, Aaron and Terveen, Loren and Zhu, Haiyi},
	year = {2020},
	pages = {1--14},
}

@article{chauhan_value_2024,
	address = {New York, NY, USA},
	title = {Value {Tensions} in {OpenStreetMap}: {Openness}, {Membership}, and {Policy} in {Online} {Communities}},
	volume = {8},
	url = {https://doi.org/10.1145/3686919},
	doi = {10.1145/3686919},
	number = {CSCW2},
	journal = {Proceedings of the ACM on Human-Computer Interaction},
	publisher = {Association for Computing Machinery},
	author = {Chauhan, Aarjav and Sarkar, Dipto and Agrawaal, Taneea S. and Soden, Robert},
	month = nov,
	year = {2024},
}

@article{hill_hidden_2021,
	title = {The {Hidden} {Costs} of {Requiring} {Accounts}: {Quasi}-{Experimental} {Evidence} {From} {Peer} {Production}},
	volume = {48},
	issn = {0093-6502},
	shorttitle = {The {Hidden} {Costs} of {Requiring} {Accounts}},
	url = {https://doi.org/10.1177/0093650220910345},
	doi = {10.1177/0093650220910345},
	language = {EN},
	number = {6},
	urldate = {2026-06-17},
	journal = {Communication Research},
	publisher = {SAGE Publications Inc},
	author = {Hill, Benjamin Mako and Shaw, Aaron},
	month = aug,
	year = {2021},
	pages = {771--795},
}

@article{horta_ribeiro_post_2025,
	title = {Post {Guidance} for {Online} {Communities}},
	volume = {9},
	url = {https://dl.acm.org/doi/10.1145/3711046},
	doi = {10.1145/3711046},
	number = {2},
	urldate = {2026-07-02},
	journal = {Proceedings of the ACM on Human-Computer Interaction},
	author = {Horta Ribeiro, Manoel and West, Robert and Lewis, Ryan and Kairam, Sanjay},
	month = may,
	year = {2025},
	pages = {CSCW148:1--CSCW148:26},
}

@article{zhou2025llms,
  title={Llms in wikipedia: Investigating how llms impact participation in knowledge communities},
  author={Zhou, Moyan and Cho, Soobin and Terveen, Loren},
  journal={arXiv preprint arXiv:2509.07819},
  year={2025}
}

@inproceedings{brooks_rise_2024,
	address = {Miami, Florida, USA},
	title = {The {Rise} of {AI}-{Generated} {Content} in {Wikipedia}},
	url = {https://aclanthology.org/2024.wikinlp-1.12/},
	doi = {10.18653/v1/2024.wikinlp-1.12},
	urldate = {2026-06-05},
	booktitle = {Proceedings of the {First} {Workshop} on {Advancing} {Natural} {Language} {Processing} for {Wikipedia}},
	publisher = {Association for Computational Linguistics},
	author = {Brooks, Creston and Eggert, Samuel and Peskoff, Denis},
	editor = {Fan, Angela and Gwadabe, Tajuddeen and Johnson, Isaac and Kaffee, Lucie-Aimée and Petroni, Fabio and van Strien, Daniel},
	month = nov,
	year = {2024},
	pages = {67--79},
}

@article{liang_quantifying_2025,
	title = {Quantifying large language model usage in scientific papers},
	volume = {9},
	copyright = {2025 The Author(s), under exclusive licence to Springer Nature Limited},
	issn = {2397-3374},
	url = {https://www.nature.com/articles/s41562-025-02273-8},
	doi = {10.1038/s41562-025-02273-8},
	language = {en},
	number = {12},
	urldate = {2026-07-02},
	journal = {Nature Human Behaviour},
	publisher = {Nature Publishing Group},
	author = {Liang, Weixin and Zhang, Yaohui and Wu, Zhengxuan and Lepp, Haley and Ji, Wenlong and Zhao, Xuandong and Cao, Hancheng and Liu, Sheng and He, Siyu and Huang, Zhi and Yang, Diyi and Potts, Christopher and Manning, Christopher D. and Zou, James},
	month = dec,
	year = {2025},
	pages = {2599--2609},
}

@misc{veselovsky_artificial_2023,
	title = {Artificial {Artificial} {Artificial} {Intelligence}: {Crowd} {Workers} {Widely} {Use} {Large} {Language} {Models} for {Text} {Production} {Tasks}},
	shorttitle = {Artificial {Artificial} {Artificial} {Intelligence}},
	url = {http://arxiv.org/abs/2306.07899},
	doi = {10.48550/arXiv.2306.07899},
	urldate = {2026-07-02},
	publisher = {arXiv},
	author = {Veselovsky, Veniamin and Horta Ribeiro, Manoel and West, Robert},
	month = jun,
	year = {2023},
	note = {arXiv:2306.07899 [cs.CL]},
}

@article{veselovsky_prevalence_2025,
	title = {Prevalence and {Prevention} of {Large} {Language} {Model} {Use} in {Crowd} {Work}},
	volume = {68},
	issn = {0001-0782},
	url = {https://dl.acm.org/doi/10.1145/3685527},
	doi = {10.1145/3685527},
	number = {3},
	urldate = {2026-07-02},
	journal = {Communications of the ACM},
	author = {Veselovsky, Veniamin and Horta Ribeiro, Manoel and Cozzolino, Philip J. and Gordon, Andrew and Rothschild, David and West, Robert},
	month = feb,
	year = {2025},
	pages = {42--47},
}

@article{burtch_consequences_2024,
	title = {The consequences of generative {AI} for online knowledge communities},
	volume = {14},
	copyright = {2024 The Author(s)},
	issn = {2045-2322},
	url = {https://www.nature.com/articles/s41598-024-61221-0},
	doi = {10.1038/s41598-024-61221-0},
	language = {en},
	number = {1},
	urldate = {2026-06-05},
	journal = {Scientific Reports},
	publisher = {Nature Publishing Group},
	author = {Burtch, Gordon and Lee, Dokyun and Chen, Zhichen},
	month = may,
	year = {2024},
	pages = {10413},
}

@article{del_rio-chanona_large_2024,
	title = {Large language models reduce public knowledge sharing on online {Q}\&{A} platforms},
	volume = {3},
	issn = {2752-6542},
	url = {https://doi.org/10.1093/pnasnexus/pgae400},
	doi = {10.1093/pnasnexus/pgae400},
	number = {9},
	urldate = {2026-07-02},
	journal = {PNAS Nexus},
	author = {del Rio-Chanona, R Maria and Laurentsyeva, Nadzeya and Wachs, Johannes},
	month = sep,
	year = {2024},
	pages = {pgae400},
}

@article{shumailov_ai_2024,
	title = {{AI} models collapse when trained on recursively generated data},
	volume = {631},
	doi = {10.1038/s41586-024-07566-y},
	number = {8022},
	journal = {Nature},
	author = {Shumailov, Ilia and Shumaylov, Zakhar and Zhao, Yiren and Papernot, Nicolas and Anderson, Ross and Gal, Yarin},
	year = {2024},
	pages = {755--759},
}

@article{vetter_endangered_2025,
	title = {An Endangered Species: How {LLMs} Threaten {Wikipedia's} Sustainability},
	pages = {1--14},
	journal = {AI \& Society},
	author = {Vetter, Matthew A. and Jiang, Jialei and McDowell, Zachary J.},
	year = {2025},
}

@inproceedings{lyu_wikipedia_2025,
	address = {New York, NY, USA},
	series = {{WWW} '25},
	title = {Wikipedia {Contributions} in the {Wake} of {ChatGPT}},
	doi = {10.1145/3701716.3715543},
	booktitle = {Companion {Proceedings} of the {ACM} on {Web} {Conference} 2025},
	publisher = {Association for Computing Machinery},
	author = {Lyu, Liang and Siderius, James and Li, Hannah and Acemoglu, Daron and Huttenlocher, Daniel and Ozdaglar, Asuman},
	year = {2025},
}

@misc{reeves_exploring_2024,
	title = {Exploring the {Impact} of {ChatGPT} on {Wikipedia} {Engagement}},
	url = {https://arxiv.org/abs/2405.10205},
	publisher = {arXiv},
	author = {Reeves, Neal and Yin, Wenjie and Simperl, Elena},
	year = {2024},
	note = {arXiv:2405.10205},
}

@inproceedings{shao_assisting_2024,
	address = {Mexico City, Mexico},
	title = {Assisting in {Writing} {Wikipedia}-like {Articles} {From} {Scratch} with {Large} {Language} {Models}},
	url = {https://aclanthology.org/2024.naacl-long.347/},
	doi = {10.18653/v1/2024.naacl-long.347},
	urldate = {2026-06-17},
	booktitle = {Proceedings of the 2024 {Conference} of the {North} {American} {Chapter} of the {Association} for {Computational} {Linguistics}: {Human} {Language} {Technologies} ({Volume} 1: {Long} {Papers})},
	publisher = {Association for Computational Linguistics},
	author = {Shao, Yijia and Jiang, Yucheng and Kanell, Theodore and Xu, Peter and Khattab, Omar and Lam, Monica},
	editor = {Duh, Kevin and Gomez, Helena and Bethard, Steven},
	month = jun,
	year = {2024},
	pages = {6252--6278},
}

@article{petroni_improving_2023,
	title = {Improving {Wikipedia} Verifiability with {AI}},
	volume = {5},
	number = {10},
	pages = {1142--1148},
	journal = {Nature Machine Intelligence},
	author = {Petroni, Fabio and Broscheit, Samuel and Piktus, Aleksandra and Lewis, Patrick and Izacard, Gautier and Hosseini, Lucas and Dwivedi-Yu, Jane and others},
	year = {2023},
}

@inproceedings{semnani_detecting_2025,
	address = {Suzhou, China},
	title = {Detecting {Corpus}-{Level} {Knowledge} {Inconsistencies} in {Wikipedia} with {Large} {Language} {Models}},
	isbn = {979-8-89176-332-6},
	url = {https://aclanthology.org/2025.emnlp-main.1765/},
	doi = {10.18653/v1/2025.emnlp-main.1765},
	urldate = {2026-06-17},
	booktitle = {Proceedings of the 2025 {Conference} on {Empirical} {Methods} in {Natural} {Language} {Processing}},
	publisher = {Association for Computational Linguistics},
	author = {Semnani, Sina and Burapacheep, Jirayu and Khatua, Arpandeep and Atchariyachanvanit, Thanawan and Wang, Zheng and Lam, Monica},
	editor = {Christodoulopoulos, Christos and Chakraborty, Tanmoy and Rose, Carolyn and Peng, Violet},
	month = nov,
	year = {2025},
	pages = {34839--34866},
}

@article{reeves_machines_2025,
	title = {Machines in the {Margins}: {A} {Systematic} {Review} of {Automated} {Content} {Generation} for {Wikipedia}},
	volume = {9},
	shorttitle = {Machines in the {Margins}},
	url = {https://dl.acm.org/doi/10.1145/3757626},
	doi = {10.1145/3757626},
	number = {7},
	urldate = {2026-07-02},
	journal = {Proceedings of the ACM on Human-Computer Interaction},
	author = {Reeves, Neal and Simperl, Elena},
	month = oct,
	year = {2025},
	pages = {CSCW445:1--CSCW445:30},
}

@misc{ashkinaze_seeing_2024,
	title = {Seeing {Like} an {AI}: {How} {LLMs} {Apply} (and {Misapply}) {Wikipedia} {Neutrality} {Norms}},
	url = {https://arxiv.org/abs/2407.04183},
	publisher = {arXiv},
	author = {Ashkinaze, Joshua and Guan, Ruijia and Kurek, Laura and Adar, Eytan and Budak, Ceren and Gilbert, Eric},
	year = {2024},
	note = {arXiv:2407.04183},
}

@misc{wiki_llm_writing,
  author = {{Wikipedia contributors}},
  title  = {{Wikipedia:Writing articles with large language models}},
  year   = {2026},
  url    = {https://en.wikipedia.org/wiki/Wikipedia:Writing_articles_with_large_language_models},
  note   = {Accessed 2026-07-05}
}

@misc{wiki_signs_ai_writing,
  author = {{Wikipedia contributors}},
  title  = {{Wikipedia:Signs of AI writing}},
  year   = {2026},
  url    = {https://en.wikipedia.org/wiki/Wikipedia:Signs_of_AI_writing},
  note   = {Accessed 2026-07-05}
}

@misc{wiki_aicleanup,
  author = {{Wikipedia contributors}},
  title  = {{Wikipedia:WikiProject AI Cleanup}},
  year   = {2026},
  url    = {https://en.wikipedia.org/wiki/Wikipedia:WikiProject_AI_Cleanup},
  note   = {Accessed 2026-07-05}
}

@misc{wikipedia_g15_2026,
  author = {{Wikipedia contributors}},
  title  = {{Wikipedia:WikiProject AI Cleanup/Policies}},
  year   = {2026},
  url    = {https://en.wikipedia.org/wiki/Wikipedia:WikiProject_AI_Cleanup/Policies},
  note   = {Accessed 2026-06-05}
}

@misc{wiki_fr_genai,
  author = {{Contributeurs de Wikipédia}},
  title  = {{Wikipédia:Intelligence artificielle générative}},
  year   = {2026},
  url    = {https://fr.wikipedia.org/wiki/Wikip%C3%A9dia:Intelligence_artificielle_g%C3%A9n%C3%A9rative},
  note   = {Accessed 2026-07-05}
}

@misc{wiki_ai_policies_by_project,
  author = {{Meta-Wiki contributors}},
  title  = {{Artificial intelligence/Policies by project}},
  year   = {2026},
  url    = {https://meta.wikimedia.org/wiki/Artificial_intelligence/Policies_by_project},
  note   = {Accessed 2026-07-05}
}

@article{froneman_failed_2026,
	title = {Failed comprehensiveness, successful minimalism: {Wikipedia}’s 3-year struggle to govern {AI}-generated content (2022–2025)},
	issn = {1435-5655},
	shorttitle = {Failed comprehensiveness, successful minimalism},
	url = {https://doi.org/10.1007/s00146-026-03046-1},
	doi = {10.1007/s00146-026-03046-1},
	language = {en},
	urldate = {2026-06-05},
	journal = {AI \& Society},
	author = {Froneman, Willemien},
	month = apr,
	year = {2026},
}

@misc{foundation_ai_2025,
	title = {In the {AI} era, {Wikipedia} has never been more valuable},
	url = {https://wikimediafoundation.org/news/2025/11/10/in-the-ai-era-wikipedia-has-never-been-more-valuable/},
	language = {en-US},
	author = {{Wikimedia Foundation}},
	month = nov,
	year = {2025},
	howpublished = {\url{https://wikimediafoundation.org/news/2025/11/10/in-the-ai-era-wikipedia-has-never-been-more-valuable/}},
	note = {Accessed 2026-06-05},
}

@inproceedings{amershi_guidelines_2019,
	address = {New York, NY, USA},
	series = {{CHI} '19},
	title = {Guidelines for {Human}-{AI} {Interaction}},
	isbn = {978-1-4503-5970-2},
	url = {https://dl.acm.org/doi/10.1145/3290605.3300233},
	doi = {10.1145/3290605.3300233},
	urldate = {2026-07-02},
	booktitle = {Proceedings of the 2019 {CHI} {Conference} on {Human} {Factors} in {Computing} {Systems}},
	publisher = {Association for Computing Machinery},
	author = {Amershi, Saleema and Weld, Dan and Vorvoreanu, Mihaela and Fourney, Adam and Nushi, Besmira and Collisson, Penny and Suh, Jina and Iqbal, Shamsi and Bennett, Paul N. and Inkpen, Kori and Teevan, Jaime and Kikin-Gil, Ruth and Horvitz, Eric},
	month = may,
	year = {2019},
	pages = {1--13},
}

@inproceedings{lai_human-ai_2022,
	address = {New York, NY, USA},
	series = {{CHI} '22},
	title = {Human-{AI} {Collaboration} via {Conditional} {Delegation}: {A} {Case} {Study} of {Content} {Moderation}},
	isbn = {978-1-4503-9157-3},
	shorttitle = {Human-{AI} {Collaboration} via {Conditional} {Delegation}},
	url = {https://dl.acm.org/doi/10.1145/3491102.3501999},
	doi = {10.1145/3491102.3501999},
	urldate = {2026-07-02},
	booktitle = {Proceedings of the 2022 {CHI} {Conference} on {Human} {Factors} in {Computing} {Systems}},
	publisher = {Association for Computing Machinery},
	author = {Lai, Vivian and Carton, Samuel and Bhatnagar, Rajat and Liao, Q. Vera and Zhang, Yunfeng and Tan, Chenhao},
	month = apr,
	year = {2022},
	pages = {1--18},
}

@article{vaccaro_when_2024,
	title = {When combinations of humans and {AI} are useful: {A} systematic review and meta-analysis},
	volume = {8},
	copyright = {2024 The Author(s)},
	issn = {2397-3374},
	shorttitle = {When combinations of humans and {AI} are useful},
	url = {https://www.nature.com/articles/s41562-024-02024-1},
	doi = {10.1038/s41562-024-02024-1},
	language = {en},
	number = {12},
	urldate = {2026-07-02},
	journal = {Nature Human Behaviour},
	publisher = {Nature Publishing Group},
	author = {Vaccaro, Michelle and Almaatouq, Abdullah and Malone, Thomas},
	month = dec,
	year = {2024},
	pages = {2293--2303},
}

@article{geiger2017operationalizing,
  title={Operationalizing conflict and cooperation between automated software agents in wikipedia: A replication and expansion of'even good bots fight'},
  author={Geiger, R Stuart and Halfaker, Aaron},
  journal={Proceedings of the ACM on human-computer interaction},
  volume={1},
  number={CSCW},
  pages={1--33},
  year={2017},
  publisher={ACM New York, NY, USA}
}

@article{braun2006using,
  title={Using thematic analysis in psychology},
  author={Braun, Virginia and Clarke, Victoria},
  journal={Qualitative research in psychology},
  volume={3},
  number={2},
  pages={77--101},
  year={2006},
  publisher={Taylor \& Francis}
}

@article{castro2010methodology,
  title={A methodology for conducting integrative mixed methods research and data analyses},
  author={Castro, Felipe Gonz{\'a}lez and Kellison, Joshua G and Boyd, Stephen J and Kopak, Albert},
  journal={Journal of mixed methods research},
  volume={4},
  number={4},
  pages={342--360},
  year={2010},
  publisher={Sage Publications Sage CA: Los Angeles, CA}
}

@misc{wikimedia_source_verification_2026,
    author = {{Wikimedia Foundation Research}},
    title = {Source Verification Prototype},
    year = {2026},
    howpublished = {Wikimedia Research Report No. 14},
    url = {https://research.wikimedia.org/report_14.html},
    note = {Accessed: 2026-09-07}
}

@inproceedings{hale2014multilinguals,
  title={Multilinguals and Wikipedia editing},
  author={Hale, Scott A},
  booktitle={Proceedings of the 2014 ACM conference on Web science},
  pages={99--108},
  year={2014}
}

@misc{wikimedia_revision_risk_2023,
    author = {{Wikimedia Foundation Research}},
    title = {Enhanced Models for Content Patrolling: Multilingual Revision Risk},
    year = {2023},
    howpublished = {Wikimedia Research Report No. 8},
    url = {https://research.wikimedia.org/report_8.html},
    note = {Accessed: 2026-09-07}
}

@misc{albon2025ai,
  author       = {Chris Albon and Leila Zia},
  title        = {Our New AI Strategy Puts Wikipedia's Humans First},
  year         = {2025},
  month        = apr,
  day          = {30},
  publisher    = {Wikimedia Foundation},
  url          = {https://wikimediafoundation.org/news/2025/04/30/our-new-ai-strategy-puts-wikipedias-humans-first/},
  note         = {Accessed: 2026-09-07}
}

@misc{albon_zia_2025_ai_editors,
  author       = {Albon, Chris and Zia, Leila},
  title        = {Artificial Intelligence for Editors},
  year         = {2025},
  month        = apr,
  howpublished = {Wikimedia Foundation, Meta-Wiki},
  url          = {https://meta.wikimedia.org/wiki/Strategy/Multigenerational/Artificial_intelligence_for_editors},
  note         = {Strategy brief published in April 2025}
}

@article{hill2020wikipedia,
  title={Wikipedia and the end of open collaboration},
  author={Hill, Benjamin Mako and Shaw, Aaron},
  journal={Wikipedia},
  volume={20},
  year={2020}
}

@article{madison2024knowledge,
  title={Knowledge Commons Past, Present, and Future},
  author={Madison, Michael J},
  journal={Lewis \& Clark L. Rev.},
  volume={28},
  pages={303},
  year={2024},
  publisher={HeinOnline}
}

@inproceedings{morgan2018evaluating,
  title={Evaluating the impact of the Wikipedia Teahouse on newcomer socialization and retention},
  author={Morgan, Jonathan T and Halfaker, Aaron},
  booktitle={Proceedings of the 14th international symposium on open collaboration},
  pages={1--7},
  year={2018}
}

@book{dourish2001action,
  title={Where the action is},
  author={Dourish, Paul},
  volume={13},
  year={2001},
  publisher={MIT press Cambridge}
}

@article{genus2018collingridge,
  title={Collingridge and the dilemma of control: Towards responsible and accountable innovation},
  author={Genus, Audley and Stirling, Andy},
  journal={Research policy},
  volume={47},
  number={1},
  pages={61--69},
  year={2018},
  publisher={Elsevier}
}

@book{berker2005domestication,
  title={Domestication of media and technology},
  author={Berker, Thomas and Hartmann, Maren and Punie, Yves and Ward, Katie},
  year={2005},
  publisher={McGraw-Hill Education (UK)}
}

\onecolumn
\appendix
\section{Event-study Specification}
\label{app:event-study}
\subsection{Matching Design}
\begin{table*}[h]
\small
\centering
\caption{Summary of the two matched designs used to compare the trajectories of WikiVault-assisted and non-WikiVault articles.}
\label{tab:quasi_design}
\begin{tabular}{p{2.9cm}p{5.9cm}p{5.2cm}}
\toprule
\textbf{Component} &
\textbf{Existing Articles} &
\textbf{WikiVault-Created Articles} \\
\midrule

\textbf{Treatment} &
First WikiVault-assisted edit to an existing article &
Article created using WikiVault \\
\cmidrule(lr){1-3}

\textbf{Event time} &
Week of first WikiVault-assisted edit (\(t=0\)) &
Week of article creation (\(t=0\)) \\
\cmidrule(lr){1-3}

\textbf{Control pool} &
Never-WikiVault articles that existed at \(t=0\) and had complete observations during \(t=-4,\ldots,-1\) &
Articles created after WikiVault's deployment that were never edited using WikiVault \\
\cmidrule(lr){1-3}

\textbf{Exact matching} &
Creation month; English Wikipedia article availability at creation &
Creation week; top macro topic; English Wikipedia article availability at creation \\
\cmidrule(lr){1-3}

\textbf{Similarity matching} &
Greedy 1:1 nearest-neighbor matching without replacement on weekly pre-treatment pageviews, indicators for any non-bot edit and any content removal, article length, and topic composition, with a caliper of \(0.75\) on \(\log(1+\text{pageviews})\) in each pre-treatment week
&

Random 1:1 selection without replacement within exact-match strata \\

\cmidrule(lr){1-3}

\textbf{Balance assessment} &
Post-matching standardized mean differences (SMDs) for matching covariates &
Exact matching on all matching characteristics \\
\cmidrule(lr){1-3}

\textbf{Observation window} &
Four weeks before and four weeks after first WikiVault use &
First 12 weeks following article creation \\
\cmidrule(lr){1-3}

\textbf{Estimation} &
Event-study difference-in-differences &
Matched-pair fixed-effects regression \\
\cmidrule(lr){1-3}

\textbf{Matched pairs} &
1,767 &
4,360 \\

\bottomrule
\end{tabular}
\end{table*}

\subsection{Covariate Balance}
\label{app:matching-balance}
\begin{table}[h]
\caption{\textbf{Covariate balance after matching.}
Absolute standardized mean differences (SMDs) between WikiVault-treated
articles and matched control articles. Continuous covariates are measured
on the $\log(1+x)$-transformed values used for matching; topic indicators
are measured on their original binary scale.
Following common practice, absolute SMDs below 0.10 indicate excellent balance, while values below 0.25 are generally considered acceptable.}
\centering

\begin{tabular}{lcccc@{\hspace{1.5em}}lc}
\toprule
& \multicolumn{4}{c}{Weeks prior to treatment}
& \multicolumn{2}{c}{Topic indicators} \\
\cmidrule(lr){2-5}
\cmidrule(lr){6-7}
Covariate & $-4$ & $-3$ & $-2$ & $-1$
& Topic & $|\mathrm{SMD}|$ \\
\midrule
Pageviews
    & 0.009 & 0.017 & 0.012 & 0.025
    &  Geography & 0.020 \\
Non-bot edits
    & 0.104 & 0.082 & 0.068 & 0.099
    & Culture & 0.080 \\
Bytes removed
    & 0.106 & 0.065 & 0.070 & 0.064
    & STEM & 0.005 \\
Article length
    & 0.008 & 0.009 & 0.010 & 0.015
    & History and Society & 0.153\\
\bottomrule
\end{tabular}

\label{tab:matching_balance}

\Description{Table showing covariate balance between WikiVault-treated and matched control articles after matching. Absolute standardized mean differences are small across pre-treatment pageviews, non-bot edits, bytes removed, and article length during each of the four weeks before treatment. Topic indicators are also generally well balanced, with the largest remaining difference for History and Society at an absolute SMD of 0.151.}
\end{table}

\begin{table}[h]
    \centering
    \caption{\textbf{Sparsity of editing outcomes before WikiVault use.}
    Entries report the range in the percentage of article-weeks with zero
    activity across the four pre-treatment weeks ($t=-4$ to $-1$).
    Each treatment group contains 1,767 articles.}
    \label{tab:outcome_sparsity}
    \small

    \begin{tabular}{lcc}
        \toprule
        Outcome
        & Matched controls
        & WikiVault-treated \\
        \midrule

        Non-bot edits
        & 98.9--99.2\%
        & 97.4--98.3\% \\

        Bytes removed
        & 99.4--99.9\%
        & 98.8--99.4\% \\

        \bottomrule
    \end{tabular}
\end{table}
\newpage
\section{Outcome Trajectories}
\vspace{-0.5em}
\begin{figure}[h!]
    \caption{Trajectories of pageviews, page length (bytes), edits, and content removal. Left panels show weekly trajectories for existing articles edited with WikiVault and their matched controls, relative to the WikiVault edit; right panels show trajectories for articles created with WikiVault and their matched controls, relative to article creation. Panels A--D show mean pageviews and page length, while Panels E--H show the percentage of articles receiving any non-bot edit or having any content removed.  Week-0 values for existing WikiVault articles in Panels E and G are shown numerically rather than on the y-axis; the non-bot editing value is 100\% because the WikiVault edit itself is included in this outcome.}
\label{fig:all_trajectories}
\includegraphics[width=0.9\linewidth, alt={Four-panel figure showing mean trajectories of pageviews and page length for matched WikiVault and control articles. For existing articles, the two groups follow similar trajectories before the WikiVault edit. At the time of the WikiVault edit, WikiVault articles increase sharply in both pageviews and page length, while control articles remain comparatively stable; these differences persist over the following four weeks. For newly created articles, WikiVault articles initially receive more pageviews, but their pageviews decline over time and generally fall below those of control articles in later weeks. WikiVault-created articles are substantially longer at creation, averaging approximately 14.6 thousand bytes compared with 8.7 thousand bytes for control articles, and remain longer throughout the 12 weeks after creation. Solid lines with circles represent WikiVault articles, dashed lines with squares represent matched controls, and vertical dashed lines mark the WikiVault edit or article creation.}]{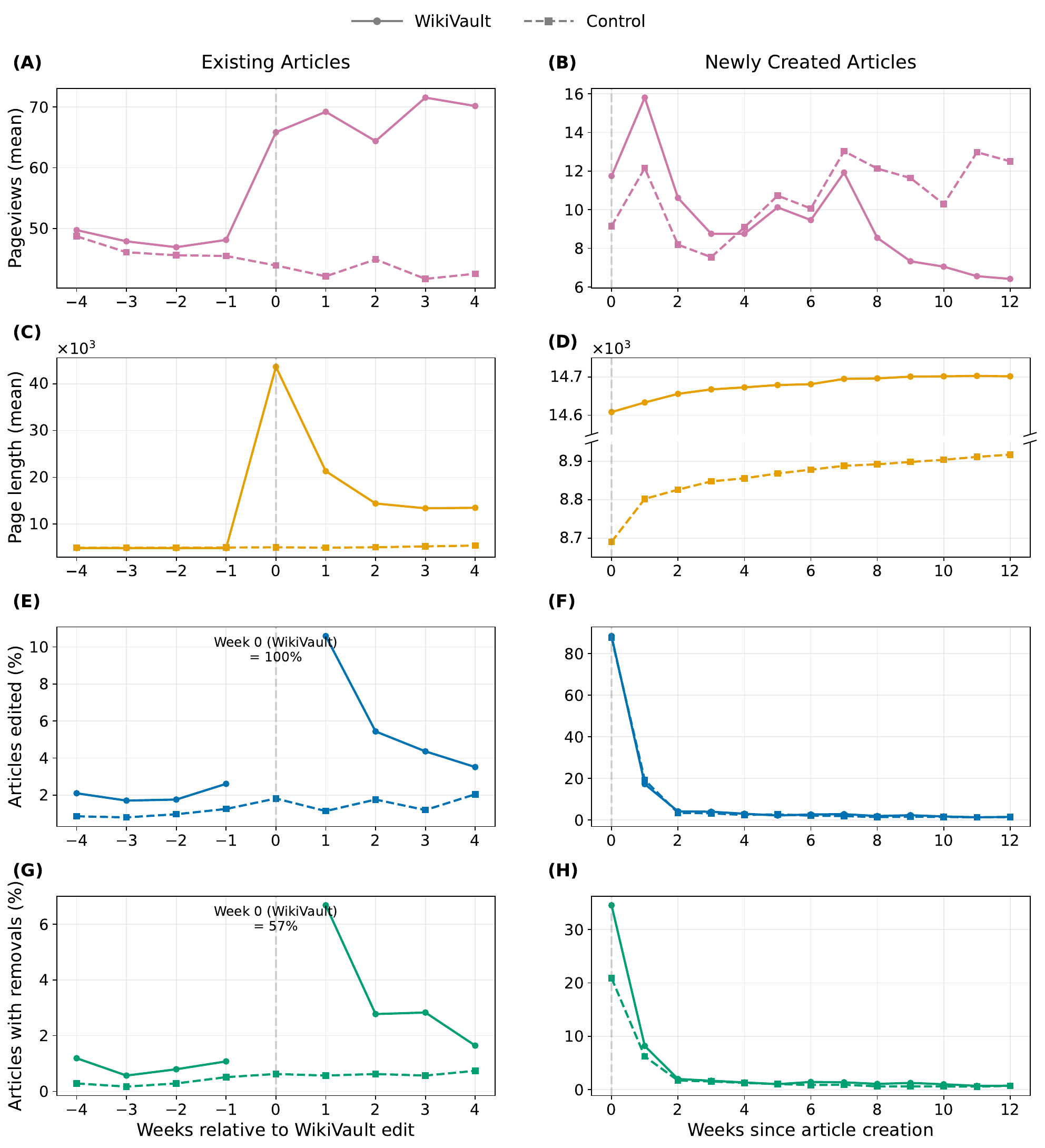}
    
\end{figure}

\end{document}